\documentclass[a4paper,fleqn,usenatbib]{mnras}
\hypersetup{draft}

\usepackage{newtxtext,newtxmath}

\usepackage[T1]{fontenc}
\usepackage{ae,aecompl}

\usepackage{graphicx}	
\usepackage{amsmath}	
\usepackage{amssymb}	
\usepackage{lineno,hyperref}
\usepackage{pdflscape}
\usepackage{multirow}

\usepackage{siunitx}
\usepackage[T1]{fontenc}
\usepackage[utf8]{inputenc}
\usepackage{gensymb}
\usepackage{wasysym}

\usepackage{esvect}

\DeclareMathOperator{\atantwo}{atan2}
\DeclareMathOperator{\sign}{sgn}

\usepackage[colorinlistoftodos]{todonotes}
\let\Oldtodo\todo
\renewcommand{\todo}[1]{\Oldtodo[inline]{#1}}

\title[Novel meteor trajectory solver]{Estimating trajectories of meteors: an observational Monte Carlo approach - I. Theory}

\author[D. Vida et al.]{
Denis Vida,$^{1,2}$\thanks{E-mail: dvida@uwo.ca}
Peter S. Gural,$^{3}$
Peter G. Brown,$^{2,4}$
Margaret Campbell-Brown,$^{2,4}$
\newauthor
Paul Wiegert$^{2,4}$
\\
$^{1}$Department of Earth Sciences, University of Western Ontario, London, Ontario, N6A 5B7, Canada\\
$^{2}$Department of Physics and Astronomy, University of Western Ontario, London, Ontario, N6A 3K7, Canada\\
$^{3}$Gural Software and Analysis LLC, Sterling, Virginia, 20164 USA\\
$^{4}$Centre for Planetary Science and Exploration, University of Western Ontario, London, Ontario, N6A 5B8, Canada\\
}

\date{Accepted 2019 November 7. Received 2019 October 28; in original form 2019 March 31}

\pubyear{2019}

\begin{document}
\label{firstpage}
\pagerange{\pageref{firstpage}--\pageref{lastpage}}
\maketitle

\begin{abstract}
It has recently been shown by \cite{egal2017challenge} that some types of existing meteor in-atmosphere trajectory estimation methods may be less accurate than others, particularly when applied to high precision optical measurements. The comparative performance of trajectory solution methods has previously only been examined for a small number of cases. Besides the radiant, orbital accuracy depends on the estimation of pre-atmosphere velocities, which have both random and systematic biases. Thus it is critical to understand the uncertainty in velocity measurement inherent to each trajectory estimation method.

In this first of a series of two papers, we introduce a novel meteor trajectory estimation method which uses the observed dynamics of meteors across stations as a global optimization function and which does not require either a theoretical or empirical flight model to solve for velocity. We also develop a 3D observational meteor trajectory simulator that uses a meteor ablation model to replicate the dynamics of meteoroid flight, as a means to validate different trajectory solvers.

We both test this new method and compare it to other methods, using synthetic meteors from three major showers spanning a wide range of velocities and geometries (Draconids, Geminids, Perseids). We determine which meteor trajectory solving algorithm performs better for: all-sky, moderate field of view, and high-precision narrow-field optical meteor detection systems. The results are presented in the second paper in this series. Finally, we give detailed equations for estimating meteor trajectories and analytically computing meteoroid orbits, and provide the Python code of the methodology as open source software.

\end{abstract}




\section{Introduction} \label{sec:introduction}

\cite{schiaparelli1871entwurf} were the first to show the connection between the orbits of meteor showers and comets \citep{romig1966scientific, hughes1982history}. This physical connection motivated development of various methods of estimating meteor trajectories, with the first reasonably precise measurements made even earlier with the pioneering work of Brandes and Benzenburg in the late 18$^{th}$ century \citep{Burke1986}. These techniques typically use optical measurements from multiple sites to estimate atmospheric meteor trajectories. \cite{gural2012solver} provides a good historical overview. 

In this work we focus on three foundational papers which provide representative descriptions of the three most common  modern meteor trajectory estimation methods. These are:
\renewcommand{\labelenumi}{\alph{enumi})}
\begin{enumerate}
    \item{the intersecting planes (IP) method as described by \cite{ceplecha1987geometric}}
    \item{the lines of sight (LoS) method by \cite{borovicka1990comparison}}
    \item{the multi-parameter fit (MPF) method of \cite{gural2012solver}}.
\end{enumerate}

The goal of any trajectory solver is to reconstruct the atmospheric trajectory of a meteor, leading ultimately to an estimate of its pre-atmospheric orbit. The trajectory is defined by a position vector (a reference position in space) and a velocity vector. To compute a reliable heliocentric orbit this should preferably be at a point before any significant deceleration of the meteoroid occurs. A common assumption is that the trajectory is a straight line, a good approximation for shorter meteors.  However, longer meteors, particularly those entering at shallow angles, may show significant deviation from a straight-line trajectory due to Earth's gravity \citep{Ceplecha1979a}. 

Existing methods usually estimate the geometry of the meteor path separately from the dynamics of the meteoroid (i.e. the time dependent characteristics of the meteor: position, velocity, acceleration). The velocity can be estimated by fitting an empirical model to the observations of time versus path length from the beginning of the meteor. \cite{gural2012solver} was the first to note that trajectories can be better constrained by fitting a meteor propagation model to both the meteor trajectory geometry and the meteoroid dynamics at the same time. This assumption makes use of the  fact that all observers should see the same dynamical behaviour of a particular meteoroid at the same point in time. A consequence of this approach is that it allows an estimate of the absolute timing offsets between stations. A further recent advance in this area is using particle filters to directly fit numerical meteor ablation models to better estimate trajectories of fireballs \citep{sansom2017analyzing}.

The original motivation for this work was earlier analysis of two station meteor data obtained by the Canadian Automated Meteor Observatory (CAMO) mirror tracking system \citep{weryk2013camo}. The system achieves an angular precision for meteor positions on the order of a few arc seconds (limited largely by the system's ability to resolve the physical spreading of the meteor itself \citep{stokan2013}), which translates to a spatial precision of a few meters. The temporal precision of the system is \SI{10}{\milli \second}. This is sufficient to discern individual fragments of fragmenting faint meteors \citep{subasinghe2016physical, vida2018canadian}. Similar to \cite{egal2017challenge}, we found that the existing methods of trajectory estimation do not always provide solutions of satisfactory quality. For example, we often found with CAMO measurements that the intersecting planes and the LoS methods produce solutions where the dynamics of the meteor do not match at different stations. The MPF method, in some cases, depending on the velocity model used, had convergence issues. This suggested that in some cases forcing the meteoroid velocity to follow a closed-form empirical model did not result in a physically consistent solution. As a result of this experience we also wanted to objectively quantify the real uncertainties and formally define the true accuracy of individually measured meteor radiants and velocities as estimated using CAMO data, and by extension other optical systems.

This series of papers attempts to answer the following question: For a given type of optical system, what is the best trajectory solver to use, and what quantitative accuracy should one typically expect? We note that this is one step in the process of defining the best estimate for a meteoroid's original heliocentric orbit. The necessary additional step is accounting for deceleration due to atmospheric drag on the earliest measured luminous point of the meteor, a topic addressed in an earlier paper \citep{vida2018modeling}.

In the following sections we discuss in detail the theory behind various methods of trajectory estimation and describe our novel Monte Carlo approach. Finally, for completeness, we summarize the equations for analytically computing meteoroid orbits from trajectory information, as previously published procedures were ambiguous in several crucial steps.

\section{Overview of trajectory solvers}

A set of line-of-sight (angle-angle) measurements of meteor positions from an individual observing station describes a fan of rays when converted into a station-fixed Cartesian coordinate system. By assuming that the position of an observer can be represented by a single point in the same coordinate system (usually at the time of the middle of the meteor's trajectory), a plane can be fit through these points \citep{ceplecha1987geometric}. By repeating the procedure for $N$ different stations, one plane for each station is obtained. The intersection of every pair of planes, $\binom N 2$ pairs in total, results in a line which describes the optimal trajectory as measured from two stations. If there are more than two trajectory lines, the average of the trajectories can be computed weighted by the squared sine of the convergence angle between every plane pair. The convergence angle is the angle between a pair of planes.

\cite{borovicka1990comparison} points out a disadvantage of the intersecting planes (IP) method: when the planes are paired using observations from multiple stations, the information about the uncertainty of individual measurements can be lost because only the whole plane is taken into consideration when intersecting it with another to define a trajectory. An outlier line-of-sight measurement can shift the whole plane in a certain direction and influence the resulting trajectory. However the fit residuals will not show the influence from the sole outlier. 

Instead of pairing planes from individual stations and producing the trajectory as a secondary product, \cite{borovicka1990comparison} proposes that one can consider every measurement of meteor position as a ray emanating from the observer in the direction of the meteor at a specific point along its linear track. Each ray is usually referred to as a line-of-sight (LoS) measurement of the meteor. The trajectory is then found as the three-dimensional line which results in the minimal distance to all measurement lines-of-sight, with the solution computed using a least squares minimization. Furthermore, \cite{borovicka1990comparison} points out that this method can compensate for Earth's rotation at each LoS observation directly during the trajectory estimation process. In the absence of this compensation, fixed observers on the non-inertial rotating surface of the Earth  perceive a virtual force (the Coriolis force) on the apparent meteor trajectory. 

Additionally, the \cite{borovicka1990comparison} method  makes  possible compensation for diurnal aberration, an effect due to the Earth's rotation that occurs because of the changing observer's perspective of the meteor with respect to distant stars. Assuming one knows the absolute time, an Earth-centred Inertial (ECI) reference frame can be adopted in which the observer's coordinates are constantly changing due to Earth's rotation, but the meteor trajectory remains linear. We use the definition of ECI coordinates where the x-axis is aligned with the mean equinox at 12:00 Terrestrial Time on January 1, 2000 (J2000).

In the original LoS paper, \cite{borovicka1990comparison} keeps the observers in the Earth-centred Earth-fixed frame (ECEF), presumably because the timing of each individual measurement (taken on a single photographic film in that era) was unknown. In contrast to the ECI system, which does not rotate with respect to the stars but the coordinates of observers on Earth's surface are changing in time, coordinates of ECEF are fixed with respect to the Earth's surface. Without correcting for the changes in observer positions, \cite{borovicka1990comparison} found the results of the IP and LoS comparable. The reason that it is not possible to account for moving observers in the intersecting planes method is that the motion of the observer and the positions of the meteor are not co-planar (unless all measurements coincide with the observer's zenith, an impossible geometry to have from two different stations). 

\begin{figure*}
  \includegraphics[width=\linewidth]{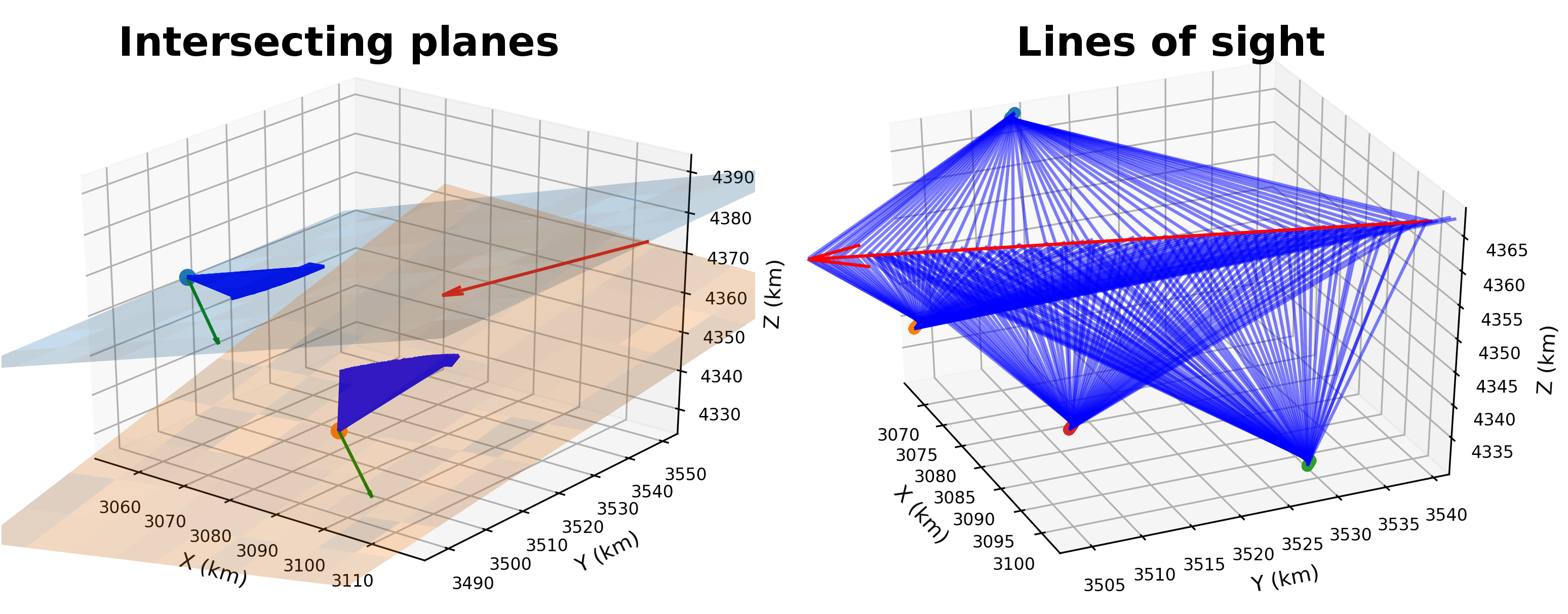}
  \caption{Left: The intersecting planes method with only the two stations having the best convergence angle shown. The planes are shown in blue and orange (semi-transparent) and are coplanar with line-of-sight observations (blue arrows emanating from stations). Note that stations are single points in the ECEF (Earth-centered Earth-fixed) frame, but here we show them in the ECI (Earth-centered inertial) frame at a fixed time. The green arrows are plane normals from each station, and the red arrow is the resulting estimated trajectory.  Right: Lines of sight method, where coordinates of all four stations are changing due to the Earth's rotation.}
  \label{fig:traj_solver_comparison}
\end{figure*}

To provide a concrete estimate of the magnitude of the diurnal aberration correction, let us consider an observer at a latitude of $\ang{45} $~N where the Earth's rotational E-W velocity is about \SI{328}{\m \per \s}. For a meteor of \SI{1}{\s} duration, the real position of the observer will change \SI{\pm 164}{\m} with respect to the average time of the trajectory determination. There is also a small effect when two observers are not at equal latitudes. A second observer at $\ang{46}$~N (\SI{\sim 120}{\kilo \metre} away) experiences a rotational velocity of \SI{322}{\m \per \s}, which causes a differential of \SI{3}{\m} between the first and the last positions of the two observers. This effect is minor if positional errors are orders of magnitude larger, but it has to be taken into account when estimating high precision trajectories where positional measurements are on the order of meters. Figure \ref{fig:traj_solver_comparison} shows a general comparison between the intersecting planes and the LoS method.

The multi-parameter fit method was first presented at the 2011 International Meteor Conference with the underlying algorithmic details described in \cite{gural2012solver}. It had been developed for the Cameras for Allsky Meteor Surveillance (CAMS) project where the full processing pipeline is described in \cite{jenniskens2011cams}. In contrast to the IP and LoS methods, which are purely geometrical, the multi-parameter fit (MPF) method uses a velocity model (i.e. dynamical information) as well. By assuming an empirical velocity model that may include deceleration terms, the MPF finds a trajectory solution (a line in 3D space) as well as the velocity and deceleration coefficients which bests describes the observed meteor's observations from all stations given the constraints of the empirical dynamical model. Because of the dynamical constraints, the method is also able to estimate relative timing offsets between camera sites.

To avoid issues of confusion with local minima in the method's cost function, an initial guess for the solution is obtained using the intersecting planes method. This guess is further refined using the LoS method - the latter modified by minimizing the angles between the measured lines of sight and the model trajectory, instead of minimizing the distances between the two. This refined guess is fed into a simplex-based non-linear equation solver where the angles between the measured lines of sight and the positions predicted by the model are minimized. This effectively ensures that all observers ``see'' the same dynamics of the meteor in time. 

Because observing systems usually do not have absolute synchronized time, the time difference between the observers must be estimated as well in the MPF. In \cite{gural2012solver}, the MPF was compared to IP and LoS using data from wide-field systems. The results showed that the radiant dispersion of meteor showers is significantly smaller if the MPF method with a constant velocity model (ie. no deceleration) is used, especially for cases with small convergence angles. The authors proposed three meteor propagation models:

	1) the constant velocity model:

\begin{equation}
    d(t) = v_0 t
\end{equation}

\noindent where $d(t)$ is the distance of the meteor at a particular point in time after the beginning point, and $v_0$ is the constant velocity of the meteor.

	2) the linear deceleration model:

\begin{equation} \label{eq:linear_deceleration_model}
    d(t)= 
    \begin{cases}
        v_0 t,                               & \text{if } t < t_0 \\
        v_0 t - \frac{1}{2} a (t - t_0)^2,   & \text{otherwise}
    \end{cases}
\end{equation}

\noindent where $t_0$ is the time when meteor begins decelerating with a constant deceleration $a$. 

	3) the empirical exponential deceleration model of \cite{whipple1957reduction}:

\begin{equation} \label{eq:exp_deceleration_model}
    d(t) = v_0 t - |a_1| e^{|a_2|t}
\end{equation}

\noindent where $a_1$ and $a_2$ are deceleration parameters.

The complexities due to the physical properties of the meteoroids and their resulting ablation behavior are not included in these models. The exponential deceleration model is the only one motivated by a physical basis, namely that the meteoroid's deceleration is proportional to the atmospheric density, following classical single-body ablation models \citep{ceplecha1998meteor}. As the atmospheric density increases exponentially with decreasing height, the velocity should follow a similar functional trend. However, the single-body assumption breaks down when a meteoroid starts fragmenting, a behaviour shown to exist for at least 90\% of meteors from high-precision observations \citep{subasinghe2016physical}, a phenomenon understood to be ubiquitous across all meteoroid masses \citep{ceplecha1998meteor, Hawkes1975}. Comparing the performance of different trajectory estimation methods even for single-body ablation has not been rigorously addressed. \cite{gural2012solver} performed simulations for a constant velocity model over an extensive range of encounter geometries and speeds. However, the comparison did not examine other functional forms of deceleration. 

In recent work by \cite{egal2017challenge} it was shown that the exponential model is difficult to fit using local cost function minimization methods since it is mathematically ill-conditioned and the associated model coefficients have linear dependencies. The authors showed the advantages of global minimization methods over local techniques. In particular, they applied the particle swarm optimization (PSO) method \citep{eberhart1995new} for fitting the exponential deceleration model and showed that it produced superior results, albeit at the expense of higher computational costs. Their work has shown that the fit works well on simulated data produced using the exponential deceleration model. In contrast, the fits were poorer when model data was created using the meteor ablation model of \cite{borovivcka2007atmospheric}. They concluded that of all methods tested (IP, LoS an MPF), the multi-parameter fit consistently produced results with the smallest residuals and good radiant solutions even for meteors with very low convergence angles ($Q_c \sim \ang{1}$). They also showed that the initial velocity estimated from all of the trajectory solvers for ablation-simulated meteoroids was not accurately determined. This suggests that a more reliable meteor propagation model is needed for the MPF in particular.

We have directly used all implementations of the three trajectory solvers including the PSO based implementation of the \cite{gural2012solver} method, to test their relative performance on high-precision CAMO data. Given the performance limitations of existing algorithms, it was decided to develop a Monte Carlo trajectory solver specifically to attempt to improve the accuracy of meteor trajectory solutions where high precision data is available. The details of this trajectory solver are given in section \ref{section:mc_trajectory}. To verify the performance of this new method and compare to the three other solvers, we also developed an observational meteor trajectory simulator. This provides synthetic measurement inputs to each solver using known solutions; details are given in section \ref{sec:trajsim}.

We emphasize that the ultimate limitation to the accuracy in the estimation of a meteoroid orbit based on observations of a meteor in the atmosphere is the amount of deceleration that occurs prior to the luminous phase. We use the term “initial velocity” for the velocity of the meteor at the moment of first detection, and ``pre-atmospheric velocity'' for the velocity before any significant deceleration has occurred (we assume this to be at a height of \SI{180}{\kilo \metre}). The difference between the initial velocity and the pre-atmospheric velocity for various types of meteoroids as measured by several typical observation systems was analyzed in \cite{vida2018modeling}. It was found that low-velocity meteors significantly decelerate (up to \SI{750}{\metre \per \second} for moderate and narrow field of view optical systems) prior to sensor detection of the visible meteor trail. The proposed correction of \cite{vida2018modeling} should be used to reconstruct the real pre-atmosphere velocity from the measured initial velocity. Establishing the latter quantity and its true uncertainty is the focus of this work.

\section{Monte Carlo trajectory estimation method} \label{section:mc_trajectory}

Our newly developed method of trajectory estimation builds on the work of \cite{gural2012solver} and expands on an earlier similar approach described in \cite{weryk2012simultaneous}. This technique uses the intersecting planes and the LoS methods to obtain a first estimate of the trajectory solution, then uses the observed angular residuals between the measurements and the fitted trajectory as a direct estimate of the uncertainty. With these estimates in hand,  Monte Carlo runs are then generated by adding Gaussian noise to the observations using the standard deviation of the angular residuals from the initial trajectory estimate and redoing the trajectory solution using noise-added data. 

This procedure gives a set of trajectories which are geometrically possible to fit within the measurement uncertainty. The lines of sight from individual stations are then projected to the trajectory line and the dynamics of the meteor as seen from every station are computed. Critically in this new technique, the best solution is chosen by comparing the observed dynamics between different stations and choosing the trajectory which has the most consistent dynamics as seen from all stations. This approach constrains the trajectory solution both geometrically and dynamically without limiting the motion to an empirical propagation/ablation model, while simultaneously keeping LoS vectors within measurement uncertainty. Note that unlike in the MPF method, the geometry and dynamics are solved separately; the dynamics is only used as an additional constraint on the geometry.

Here we provide detailed formulations of all the equations used by this trajectory solver, with the exception of well known mathematical and numerical methods. The equations are given in a way that would make their computer implementation unambiguous and thus may slightly deviate from standard mathematical notation. Where the function for the four-quadrant inverse tangent is used, we assume that the order of arguments is $atan2(y, x)$, as in e.g. C, FORTRAN, Python, and MATLAB. This differs from e.g. Mathematica and MS Excel whose implementations have the two arguments reversed. $\bmod$ is the modulo operator, the integer division remainder operation. The Python implementation of both the simulator and the solver is open source and publicly available at \url{https://github.com/wmpg/WesternMeteorPyLib}.

\subsection{Inputs and conversions to rectangular coordinates}

For every station $k \in \{1, ... N_{stations}\}$, we have measurements $j \in \{1, ..., N_{meas(k)} \}$, producing inputs to the trajectory solver:

\begin{enumerate}
    \item Relative time $t_{kj}$ in seconds of each measurement from every station, relative to the reference Julian date ${JD}_{ref}$.
    \item Angular measurements of meteor positions in the horizontal coordinate system: azimuth measured eastward from the north $A_{kj}$, and altitude above the horizon $a_{kj}$ for the epoch of date from each station. Equivalently, right ascension $\alpha$ and declination $\delta$ may be used which can be converted to azimuth and altitude using equations given in Appendix \ref{appendix:radec2altaz}. If the equatorial coordinates are given in the J2000 epoch, care must be taken to first precess them to the epoch of date (see Appendix \ref{appendix:precession_eq}). The epoch of date is assumed to be at $JD_{ref}$.
    \item Geographical coordinates of every station: geodetic latitude $\varphi_k$, longitude $\lambda_k$, and height above a WGS84 Earth ellipsoid $h_k$ (note that this height is not the same as the Mean Sea Level height reported by Google Earth and newer GPS devices - the difference can be up to 100 meters).
\end{enumerate}

The first step in the process is to compute the Julian date of every individual measurement:

\begin{equation}
    {JD}_{kj} = {JD}_{ref} + t_{kj}/86400
\end{equation}

These times get updated in the second stage of the iteration when the trajectory is recomputed after the timing offset estimation. Next, measurements are converted to equatorial coordinates for the epoch of date using equations given in Appendix \ref{appendix:altaz2radec}. Two sets of equatorial coordinates are obtained: the first assumes the stations are fixed at ${JD}_{ref}$ and are used for the intersecting planes method while the second one takes into account the movement of the stations at each measurement time step. Thus, When computing values for the intersecting plane method, the ${JD}_{ref}$ reference time should be used for all measurement points. When computing values for the lines of sight method, the Julian date ${JD}_{kj}$ of the individual measurements should be used. The measurements are then converted to Cartesian unit vectors using equation \ref{eq:radec2cartesian}. These vectors define the direction of the line of sight from a given station at each measurement point in time.

\begin{eqnarray}  \label{eq:radec2cartesian}
\begin{aligned}
    \xi = \cos \delta \cos \alpha\\
    \eta = \cos \delta \sin \alpha\\
    \zeta = \sin \delta
\end{aligned}
\end{eqnarray}

The geographical positions of the stations are converted to Earth-Centered Inertial (ECI) coordinates relative to the center of the Earth using equations given in Appendix \ref{appendix:geo2eci}. Two sets of coordinates are calculated: $X_{kj}$, $Y_{kj}$, $Z_{kj}$ for the position of each station at every point in time ${JD}_{kj}$, and $X_{k}', Y_{k}', Z_{k}'$ for stations fixed at ${JD}_{ref}$. ECI coordinates fixed at ${JD}_{ref}$ are needed for the intersecting planes method, as this method implicitly assumes that the station is a point and its coordinates cannot move in time.

\subsection{Plane fits}
 
The best fit plane for observations from one station can be defined as:

\begin{equation}
    a \vv{x} + b \vv{y} + d = -\vv{z}
\end{equation}

\noindent where $\vv{x}, \vv{y}, \vv{z}$ are data vectors containing Cartesian unit vectors of direction $\xi, \eta, \zeta$, and a zero, which represents the position of the station, taken to be the origin of the direction vector's coordinate system:

\begin{equation}
\begin{aligned}
    \vv{x} = [0, \xi_{k1}, ..., \xi_{k N_{meas(k)}}] \\
    \vv{y} = [0, \eta_{k1}, ..., \eta_{k N_{meas(k)}}] \\
    \vv{z} = [0, \zeta_{k1}, ..., \zeta_{k N_{meas(k)}}]
\end{aligned}
\end{equation}

The problem can be written in data matrix form as:

\begin{equation}
\begin{aligned}
    \begin{bmatrix}
        x_{1} & y_{1} & 1 \\
        x_{2} & y_{2} & 1 \\
               & ...    &   \\
        x_{n} & y_{n} & 1
    \end{bmatrix}
    \begin{bmatrix}
        a \\
        b \\
        d
    \end{bmatrix}
    = -
    \begin{bmatrix}
        z_{1} \\
        z_{2} \\
        ...    \\
        z_{n}
    \end{bmatrix}
\end{aligned}
\end{equation}

If we take the data matrix and pre-multiply both sides of the equation by its transpose, and invert to solve for the unknowns, we perform the equivalent of a linear least squares fit. One should normalize the points to be relative to their mean, $\bar{x}, \bar{y}, \bar{z}$, in which case $d$ can be excluded and one dimension can be dropped. Thus, the matrix equation solution can be written as:

\begin{equation}
\begin{aligned}
    \begin{bmatrix}
        a \\
        b
    \end{bmatrix}
    = -
    \begin{bmatrix}
        \sum_{i=1}^n (x_i - \bar{x})^2              & \sum_{i=1}^n (x_i - \bar{x})(y_i - \bar{y}) \\
        \sum_{i=1}^n (x_i - \bar{x})(y_i - \bar{y}) & \sum_{i=1}^n (y_i - \bar{y})^2
    \end{bmatrix}^{-1} \\
    \times \begin{bmatrix}
        \sum_{i=1}^n (x_i - \bar{x})(z_i - \bar{z}) \\
        \sum_{i=1}^n (y_i - \bar{y})(z_i - \bar{z}) \\
    \end{bmatrix}
\end{aligned}
\end{equation}

After solving the matrix, the direction normal to the fit plane is:

\begin{equation}
    \vv{n} = [a, b, 1]^T
\end{equation}

\subsection{Plane intersections}

We now consider planes in point-normal form. After finding the unit plane normal $\hat{n}_k$ for observations from every station, we make use of the additional constraint that each normal vector must go through the position of the station in ECI coordinates $(X_{k}', Y_{k}', Z_{k}')$. For $N$ stations, there is a total of $\binom{N}{2}$ combinations of different plane intersections. Although \cite{ceplecha1987geometric} shows how to compute the weighted average trajectory for all combinations of planes, we follow the approach of \cite{gural2012solver}, where only the solution with the pair of planes that have the highest convergence angle is taken. This solution is usually satisfactory to estimate the initial estimate of the trajectory for the lines of sight method which is then refined numerically.

For every pair of planes we have their normals, $\hat{n}_A$ and $\hat{n}_B$, and position vectors for every station, $\vv{p_A} = [X_{A}', Y_{A}', Z_{A}']$ and $\vv{p_B} = [X_{B}', Y_{B}', Z_{B}']$. The convergence angle $Q_{AB}$ between the two planes is:

\begin{equation}
    \cos Q_{AB} = \hat{n}_A \cdot \hat{n}_B
\end{equation}

The apparent radiant unit vector based on these two stations is:

\begin{equation}
\begin{aligned}
    \vv{R} = \hat{n}_A \times \hat{n}_B \\
    \hat{R} = \frac{\vv{R}}{| \vv{R} | }
\end{aligned}
\end{equation}

We also make sure that the radiant vector is pointing in the correct direction:

\begin{equation}
\begin{aligned}
    \hat{R} =
    \begin{cases}
        -\hat{R}, & \text{if } [\xi_{A1}, \eta_{A1}, \zeta_{A1}] \cdot \hat{R} <  [\xi_{An}, \eta_{An}, \zeta_{An}] \cdot \hat{R} \\
        \hat{R},  & \text{otherwise}
    \end{cases}
\end{aligned}
\end{equation}

where $[\xi_{A1}, \eta_{A1}, \zeta_{A1}]$ is the vector pointing to the first observed point on the meteor trajectory from station A, and $[\xi_{An}, \eta_{An}, \zeta_{An}]$ is the vector pointing to the last observed point from station A. This condition follows from the fact that the radiant is always closer to the first observed point.

The equatorial coordinates of the radiant are given by:

\begin{equation}
\begin{aligned}
    \delta = \arcsin \hat{R}_z \\
    \alpha = \atantwo ( \hat{R}_y, \hat{R}_x ) \bmod 2 \pi
\end{aligned}
\end{equation}

\noindent where the $\bmod 2 \pi$ operation wraps the right ascension to the $[0, 2 \pi]$ range.

The intersection of the planes from each station forming the radiant line in three-dimensional space is now known and unit vectors from each station to the closest point on the radiant line to the respective station can be calculated as:

\begin{equation} \label{eq:w_vect}
\begin{aligned}
    \vv{w} = \hat{R} \times \hat{n} \\
    \hat{w} = \frac{\vv{w}}{ | \vv{w} | } \\
    \hat{w} =
    \begin{cases}
        -\hat{w}, & \text{if } \hat{w} \cdot [\xi_{1}, \eta_{1}, \zeta_{1}] < 0 \\
        \hat{w}, & \text{otherwise}
    \end{cases}
\end{aligned}
\end{equation}

The last equation ensures the vector is pointing from the station towards the radiant line. These vectors, $\hat{w}_{A}$ and $\hat{w}_{B}$, are calculated for both stations.

The range vectors from each station to the radiant line can be found as:

\begin{equation}
\begin{aligned}
    \Delta \vv{p} = \vv{p_A} - \vv{p_b} \\
    \cos \omega = \hat{w}_{A} \cdot \hat{w}_{B} \\
    \vv{r_A} = \frac{\cos \omega (\Delta \vv{p} \cdot \hat{w}_{B}) - \Delta \vv{p} \cdot \hat{w}_{A}}{1 - \cos^2 \omega} \hat{w}_{A} \\
    \vv{r_B} = \frac{\Delta \vv{p} \cdot \hat{w}_{B} - \cos \omega (\Delta \vv{p} \cdot \hat{w}_{A})}{1 - \cos^2 \omega} \hat{w}_{B} \\
\end{aligned}
\end{equation}
 
where $\vv{r_A}$ and $\vv{r_B}$ are vectors pointing from the stations to the respective point on the radiant line closest in range to the station.

The ECI coordinates of the position portion of the state vector are calculated by adding the ECI position of one of the stations to the appropriate range vector. We choose the station A:

\begin{equation}
    \vv{S} = \vv{p_A} + \vv{r_A}
\end{equation}

The trajectory solution from these two stations alone is thus represented by the apparent radiant unit vector $\hat{R}$ and the reference position vector $\vv{S}$.

For the case with more than two stations we also compute weights $W_k$ for every station $k$ as:

\begin{equation} \label{eq:weight}
\begin{aligned}
    P_a = \arccos \left ( \hat{R} \cdot \hat{w_k} \right) \\
    W_k = \sin^2 P_a
\end{aligned}
\end{equation}

\noindent where $w_k$ is computed from equation \ref{eq:w_vect}, and $P_a$ is the perspective angle of the trajectory, namely the angle made between the observer, the state vector, and the radiant line. In this approach the station which observes the meteor closest to perpendicular to the trajectory is given the highest weight, while stations observing the meteor ``head on'' have the lowest weights. If the perspective angle is low, small errors in meteor position measurement will propagate into large errors on the trajectory when they get projected, thus the weight of those observations needs to be reduced. The weights are kept at unity if only 2 stations are used in the solution. The $\sin^2$ weighting scheme follows \cite{ceplecha1987geometric}, with the difference of using the perspective angle instead of the convergence angle. The weighting is only used for the lines of sight method described below.

\subsection{Line of sight method}

After pairing all planes and finding the solution with the best convergence angle, the resulting vectors $\hat{R}$ and $\vv{S}$ are taken as the starting solution for the line of sight method. This method seeks to find a radiant line (a line in 3D space) that minimizes the angular differences between all observation sight lines and the radiant line. 

Let $\vv{d_{{obs}_{kj}}} = [\xi_{kj}, \eta_{kj}, \zeta_{kj}]$ be the direction vector of every measurement from station $k$, and $\vv{d_{{mod}_{kj}}}$ be the direction of the modelled radiant line as seen from that station. The trajectory solution is then $\hat{R}$ and $\vv{S}$ for which:

\begin{equation}
    min \frac{ \sum_{k=1}^{N_{stations}} \sum_{j=1}^{N_{meas(k)}} W_k \angle ( \hat{d}_{{obs}_{kj}}, \hat{d}_{{mod}_{kj}} )} {\sum_{k=1}^{N_{stations}} W_k}
\end{equation}

\noindent This sum is minimized numerically using the Nelder-Mead method. $\hat{d}_{{mod}_{kj}}$ can be calculated using:

\begin{equation}
\begin{aligned}
    \vv{d_{{mod}_{kj}}} = \vv{T_{kj}'} - \vv{p_{kj}} \\
    \hat{d}_{{mod}_{kj}} = \frac{ \vv{d_{{mod}_{kj}}} }{ | \vv{d_{{mod}_{kj}}} | }
\end{aligned}
\end{equation}

\noindent where $\vv{T_{kj}'}$ is the gravity-corrected point on the radiant line which is the closest to the measured line of sight, and $\vv{p_{kj}}$ are the ECI coordinates of station $k$ at time $j$. $\vv{T_{kj}'}$ can be computed as:

\begin{equation} \label{eq:tkj}
\begin{aligned}
    \vv{T_{kj}'} = \vv{T_{kj}} - \Delta h (t_{kj}) \frac{\vv{T_{kj}}}{|\vv{T_{kj}}|}
\end{aligned}
\end{equation}

\noindent where $\vv{T_{kj}}$ is a point on the radiant line which is the closest to the measured line of sight that can be computed using equations given in Appendix \ref{appendix:3d_line_dist}. $\Delta h$ is the height drop due to gravity computed using the equations in Appendix \ref{appendix:gravity_drop}; adding this term effectively simulates the curvature of the trajectory due to gravity. $t_{kj}$ here is the time the meteor is at point $j$ as seen from station $k$ relative to $JD_{ref}$.

The angle between the closest point on the 3D radiant line and the observed line of sight is calculated as (note that unit vectors must be used):

\begin{equation} \label{eq:angular_residuals}
\begin{aligned}
    \angle ( \hat{d}_{{obs}_{kj}}, \hat{d}_{{mod}_{kj}} ) = \arccos \left ( \hat{d}_{{obs}_{kj}} \cdot \hat{d}_{{mod}_{kj}} \right )
\end{aligned}
\end{equation}

\subsection{Computing meteor length, velocity and lag}

Once a trajectory solution is found, the location of the estimated reference state vector position $\vv{S}$ along the radiant line is moved to the beginning of the meteor. This is done by setting $\vv{S}$ to the ECI coordinates on the radiant line with the largest observed height, implicitly assuming that a meteor is always descending downward (not necessarily true for Earth-grazers). 

The length along the track is found by projecting the observations on the radiant line using the equations given in Appendix \ref{appendix:3d_line_dist}, producing $\vv{d_{{mod}_{kj}}}$. The meteor length is defined as the distance from the reference state vector position $\vv{S}$ to every projected measurement ray along the radiant line:

\begin{equation}
\begin{aligned}
    l_{kj} = | \vv{d_{{mod}_{kj}}} - \vv{S} |
\end{aligned}
\end{equation}

The time variation of velocity defines deceleration, but since it is the second derivative of the length versus time, deceleration itself tends to have large point-to-point variances. As a proxy for overall deceleration, we use lag. Following \cite{subasinghe2017luminous}, we define lag as ``the distance that the meteoroid falls behind an object with a constant velocity that is equal to the initial meteoroid velocity''. In that work, the authors use the first half of the meteor's trajectory to estimate the initial velocity. The limitation of this approach is that the time offsets between observations from different stations can cause errors if all observations from all sites are simultaneously used for velocity estimation. Thus, the time offsets have to be estimated first.

\subsection{Estimating timing offsets and the initial velocity} \label{subsec:timing_offset_and_v0}

To estimate timing offsets we use the fact that the computed length is insensitive to offsets in time. The timing offset estimation is performed by using the station that first recorded the meteor as the station with reference time for all other stations, i.e. it has absolute time ($\Delta t = 0$). The time offsets for all stations are then numerically estimated by minimizing the sum of time differences for all combinations of station pairs. The minimization cost function $f_{\Delta t}$ is defined as:

\begin{equation} \label{eq:timing_residuals}
\begin{aligned}
    f_{\Delta t} = \frac{t_{sum}}{ W_{sum} c_{sum} } \\
    t_{sum} = \sum_{k=1}^{N_{stations}} \sum_{r=1}^{N_{stations}} \sum_{j=1}^{N_{meas(r)}} W_k W_r \left ( t_k (l_{rj}) - t_{rj} \right )^2 \\
    W_{sum} = \sum_{k=1}^{N_{stations}} \sum_{r=1}^{N_{stations}} W_k W_r \\
    c_{sum} = \sum_{k=1}^{N_{stations}} \sum_{r=1}^{N_{stations}} {N_{overlap}} \\
\end{aligned}
\end{equation}

\noindent where $k$ is the station index, $r$ the index of all other stations (iterations where $k = r$ are skipped), $t_{rj}$ is the time from station $r$, and $t_k(l_{rj})$ is the time from station $k$ at length from station $r$. $t_k(l_{rj})$ is obtained by linear interpolation of time vs. length. $W_k$ and $W_r$ are weights for the respective stations as defined in Eq. \ref{eq:weight}, and $N_{overlap}$ is the number of points that overlap in length between stations $k$ and $r$. Thus, only overlapping segments of the meteor path for stations $k$ and $r$ are used. This requirement is the main limitation of the method: for the approach to work an overlap of at least 4 points between stations is needed. If there is no overlap (e.g. one station observed only the beginning, and the other only the end of a meteor) the approach will not work and one has to assume a velocity model. For those cases we found the MPF method of \cite{gural2012solver} worked well. 

This approach of estimating time offsets is not sensitive to the functional form of the deceleration, it relies on that fact that a truly accurate trajectory solution must show the same dynamics from all stations. If the observed dynamics differ, that indicates the trajectory was not well estimated. This is the central foundation  of our novel approach. 

After an initial estimate is made of the timing offsets, the entire trajectory solution is repeated with updated timing offsets. ${JD}_{ref}$ is shifted to correspond to the new value of t = 0. Because the state vector $\vv{S}$ is kept at the beginning of the meteor, this means that the position of the meteor at time ${JD}_{ref}$ corresponds to $\vv{S}$.

The initial velocity is then estimated by progressively fitting a line to the solution time vs. length. This is done starting from the first 25\% of points from all stations (at least 4 points for short events) up to 80\% of all points. The best estimate of the initial velocity is the fit with the smallest standard deviation.

This modification mitigates the influence of deceleration on the initial velocity estimate, although at best it is the average velocity of the first 25\% of the trajectory. In practice we found that this approach works well. This approach was adopted because the standard deviation of the fit done on the first quarter of the trajectory is usually high due to the measurement uncertainty as meteors tend to be faint at the beginning of the trail and thus the initial velocity may be uncertain as well. As more points get included, the standard deviation tends to go down, but it will rise again if significant deceleration is present. The approach is thus a balance between choosing a fit that trades the effects of measurement uncertainty and deceleration.

\begin{figure}
  \includegraphics[width=\linewidth]{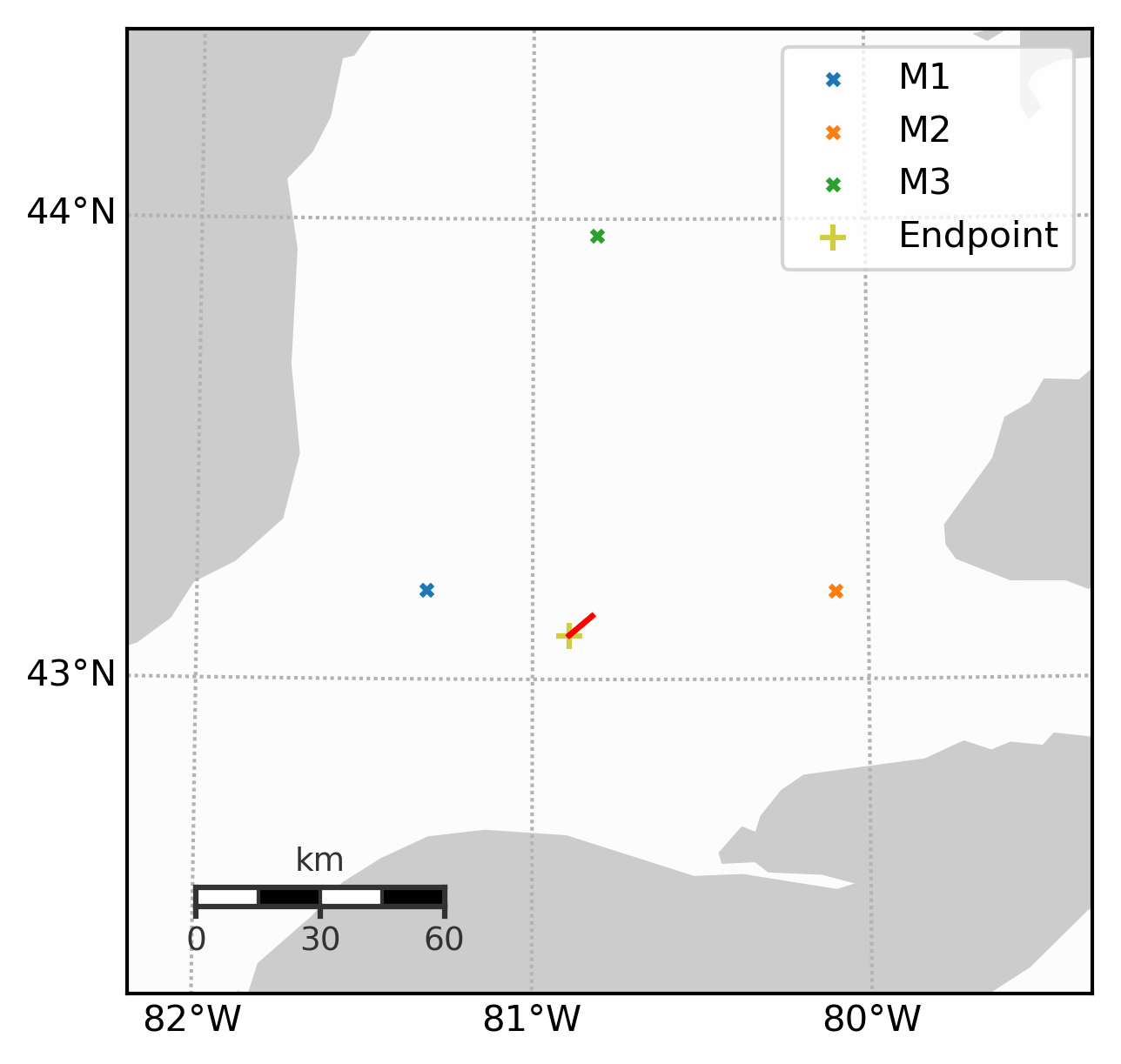}
  \caption{Map of the hypothetical moderate FOV network and the simulated Draconid of mass \SI{6.45d-05}{\kilogram}, density \SI{211}{\kilogram \per \cubic \metre} and initial velocity of \SI{23.7}{\kilo \metre \per \second}. The meteor had an entry angle of \ang{65}. Perspective angles for stations M1, M2 and M3 were \ang{19}, \ang{53} and \ang{61} respectively. The red line represents the ground track of the meteor.}
  \label{fig:sim_draconid_map}
\end{figure}

To demonstrate the accuracy of the method we have simulated a Draconid as it would be observed by a hypothetical network in Southern Ontario consisting of three stations with fields of view of $\ang{64} \times \ang{48}$ which form an equilateral triangle with sides of \SI{100}{\kilo \metre} and observe the same volume of the sky (maximum overlap at height of \SI{100}{\kilo \metre}, see the second paper for more simulation details). The accuracy of measurements was $\sigma = 0.5$ arc minutes. 

\begin{figure*}
  \includegraphics[width=\linewidth]{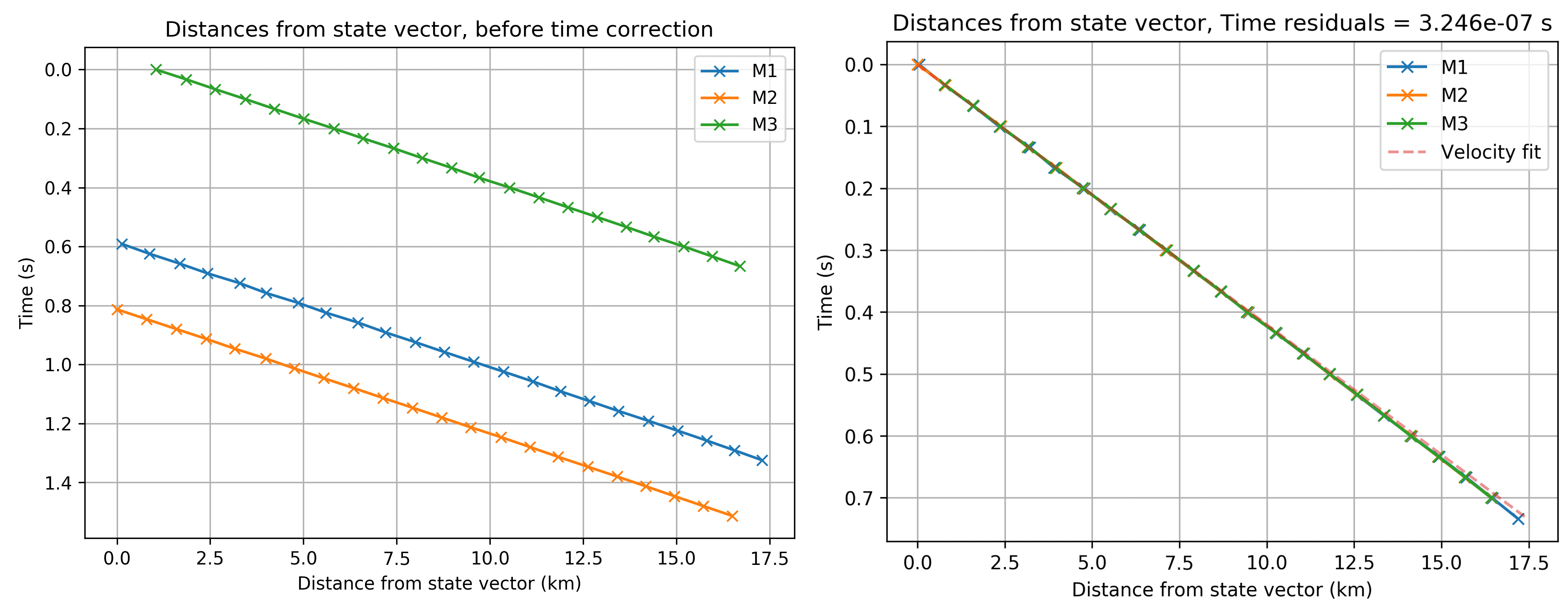}
  \caption{Left: Time vs. length before correction. Right: After time offset estimation, all curves are one on top of the other. The cited residual is the average residuals between all lines in seconds.}
  \label{fig:time_vs_length_comparison}
\end{figure*}

Figure \ref{fig:sim_draconid_map} shows the map of these model stations and the trail of the meteor. The left inset of figure \ref{fig:time_vs_length_comparison} shows time vs. length prior to the timing correction. One can see that all observations show the same trend (i.e. dynamics), but they are only offset in time. The right inset shows the lengths after estimating timing offsets and the final fitted initial velocity. Note that the observations start deviating slightly from the fitted velocity line at the end, indicating significant deceleration. 

\begin{figure}
  \includegraphics[width=\linewidth]{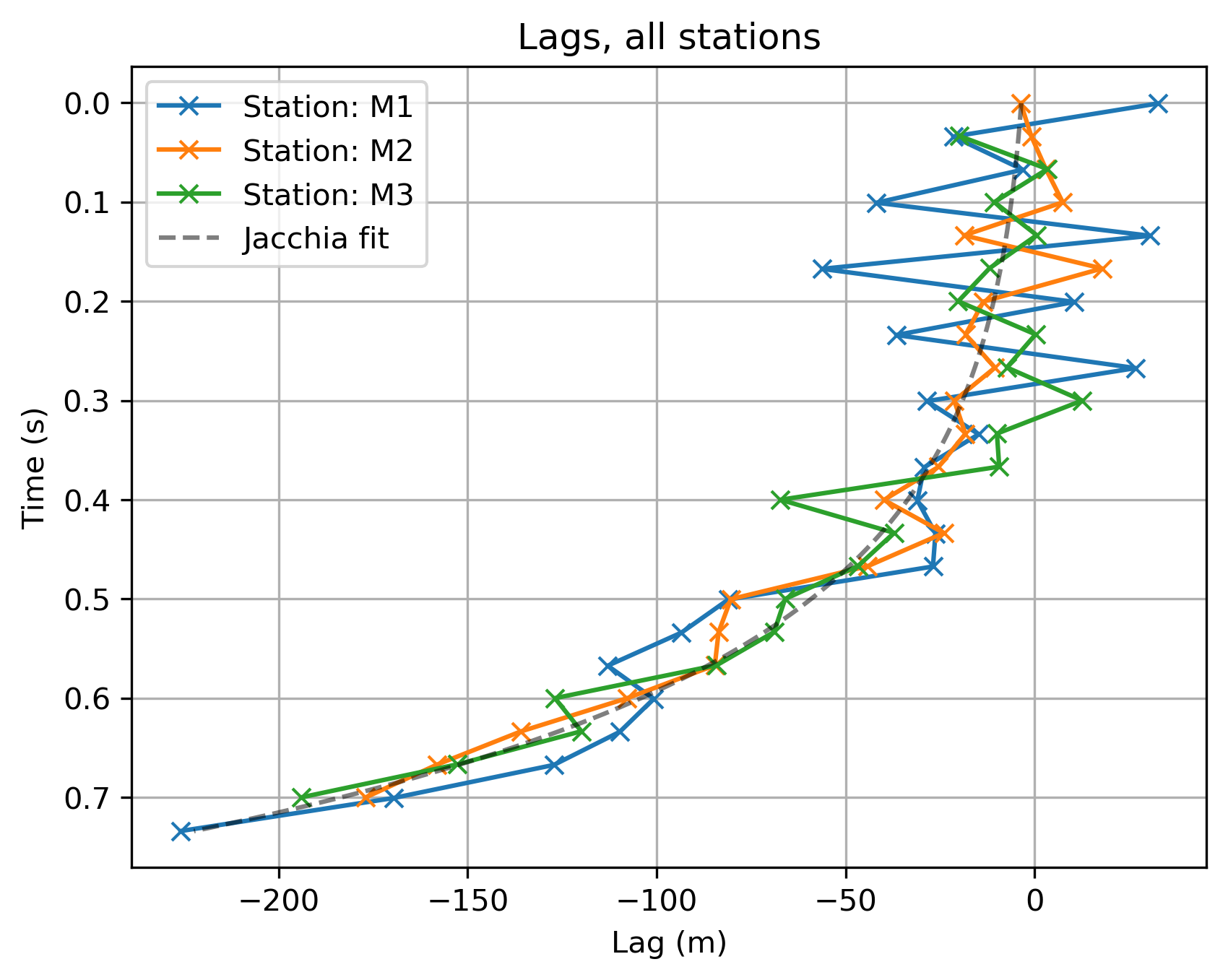}
  \caption{A lag of a simulated Draconid observed by a moderate FOV system from 3 stations. "Jacchia fit" is a fit of equation \ref{eq:exp_deceleration_model} to the computed lag.}
  \label{fig:lag_example}
\end{figure}

The effect is more visible in figure \ref{fig:lag_example} which shows the computed lag. Ideally, the lag would remain zero (a vertical line) until the meteor starts decelerating, and that straight portion would be used for initial velocity estimation. This may not always be the case if the deceleration started prior to detection, as shown in the aforementioned figure. In that case, the initial velocity will be underestimated and ablation modelling is needed to reconstruct the true initial velocity \citep{vida2018modeling}. Also, notice the larger scatter in lag and fit residuals (figure \ref{fig:angular_residuals}) from station M1 due to the low perspective angle of only \ang{19}. The perspective angles of the other two stations M2 and M3 are \ang{53} and \ang{61} respectively.

Finally, after the reference state vector, the apparent radiant, and the initial velocity are known, the orbit is computed using equations given in Appendix \ref{appendix:orbit}.

\subsection{Refining the trajectory solution - a Monte Carlo approach} \label{subsec:monte_carlo_solver}

With a nominal trajectory solution now available, the next goal is to define uncertainties in the solution and further optimize the solution using time vs. length consistency as the cost function metric.

After estimating the initial ``best'' solution as described above, the angular residuals of observations from all stations relative to this solution are computed using equation \ref{eq:angular_residuals}, as well as the value of the root-mean-square deviation (RMSD). We assume that the computed RMSD represents the standard deviation of the real (random) measurement uncertainty of individual stations. 

\begin{figure}
  \includegraphics[width=\linewidth]{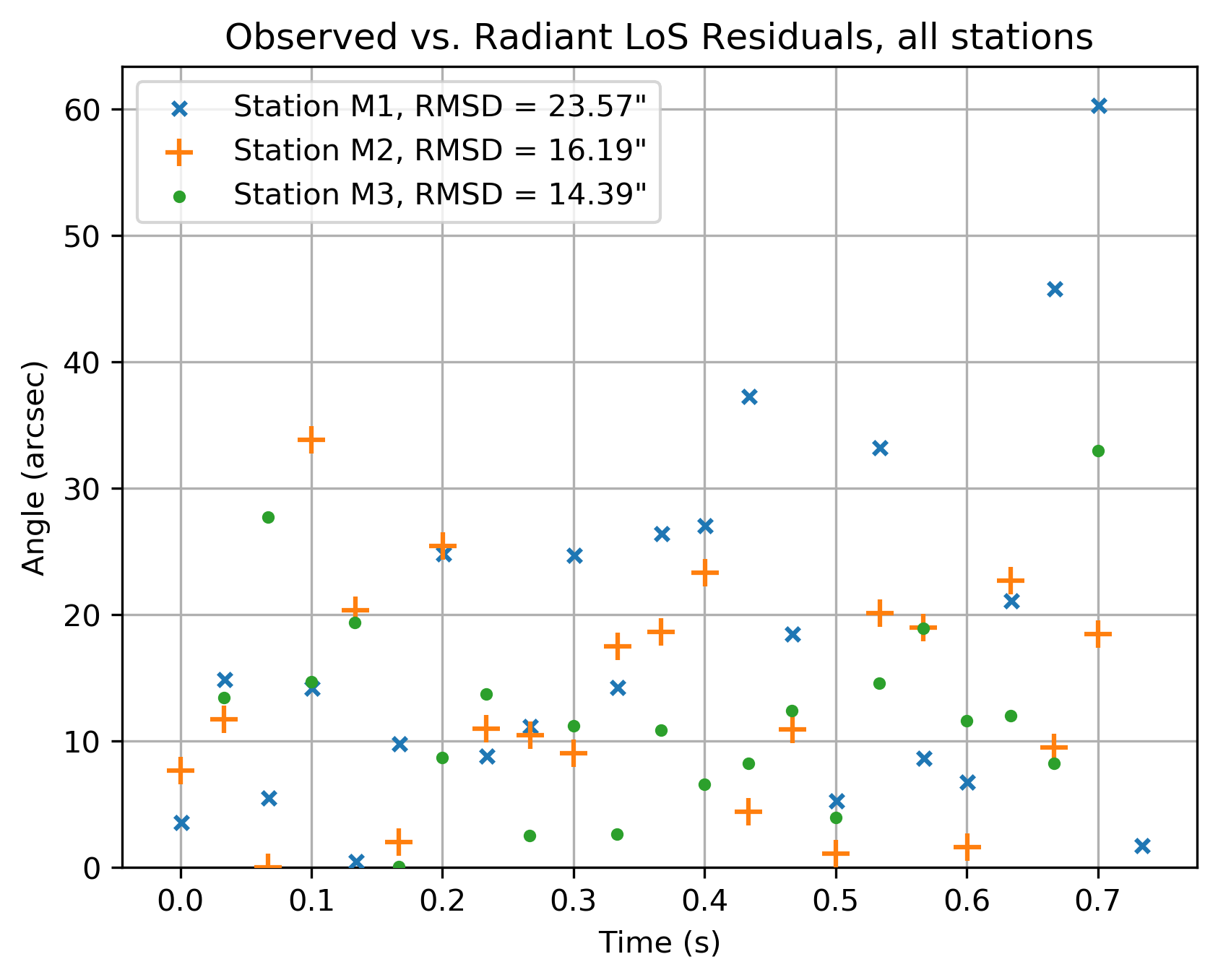}
  \caption{Angular residuals of a simulated Draconid. RMSD is the root-mean-square deviation in arc seconds.}
  \label{fig:angular_residuals}
\end{figure}

Figure \ref{fig:angular_residuals} shows the computed angular residuals for the example meteor in figure \ref{fig:lag_example}. Note that station M1 has the highest RMSD, again due to its low perspective angle. In this case, a low station weight will prevent these measurements from significantly influencing the trajectory solution.

Next, Gaussian noise is added to the original measurements from every station (using equation \ref{eq:adding_gaussian_noise}), with an a standard deviation estimated from the measured station residuals. The entire trajectory is then recomputed from the beginning and a new positional state vector, radiant, velocity and orbit are computed using the noise-added data. This procedure is repeated hundreds of times with randomized noise injected into every run. 

The best solution is chosen as the one with the smallest value of the $f_{\Delta t}$ function (equation \ref{eq:timing_residuals}). This solution is the one where the most consistent dynamics of a meteor have been observed across all stations and which is simultaneously consistent within measurement uncertainty from all stations. This produces the best dynamical solution within the geometrical uncertainty.
 
In many cases when the geometry is good and the measurements are reasonably precise, the Monte Carlo refinement will not provide additional improvement beyond the initial solution. The comparison of the performance of the Monte Carlo solver to other trajectory solvers on simulated data is given in the second paper in this series.

The measurement uncertainty of every estimated parameter (including the orbital parameters) is computed using the subset of Monte Carlo trajectories which have values of the $f_{\Delta t}$ function smaller than that of the initial purely geometrical solution. If all solutions were to be used for uncertainty estimation, then the uncertainties would be completely driven by geometric uncertainties. This culling removes all solutions which have worse fits to the dynamics between stations than the geometrical solution, thus the dynamical constraints are included. Note that this approach does not estimate possible systematic errors arising from the astrometric calibration and position picks, which are system-dependent and should be handled separately.

\begin{figure}
  \includegraphics[width=\linewidth]{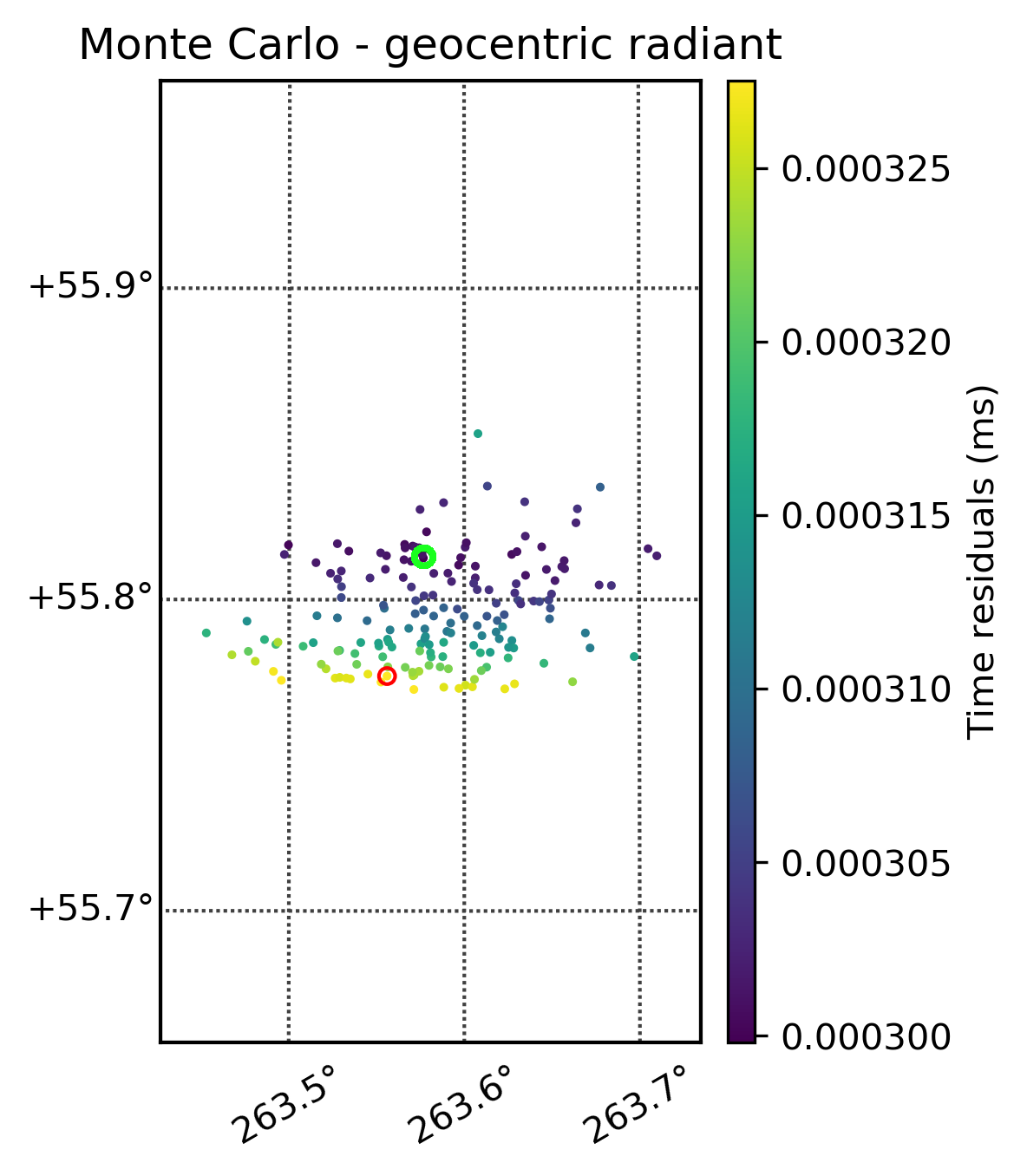}
  \caption{Spread in the geocentric radiant of the model Draconid. The square root of the timing residual $f_{\Delta t}$ is colour coded. The red circle marks the position of the initial solution $f_{\Delta t} = 0.000326$, and the green circle marks the position of the best solution $f_{\Delta t} = 0.000300$.}
  \label{fig:mc_timing_res}
\end{figure}

\begin{figure}
  \includegraphics[width=\linewidth]{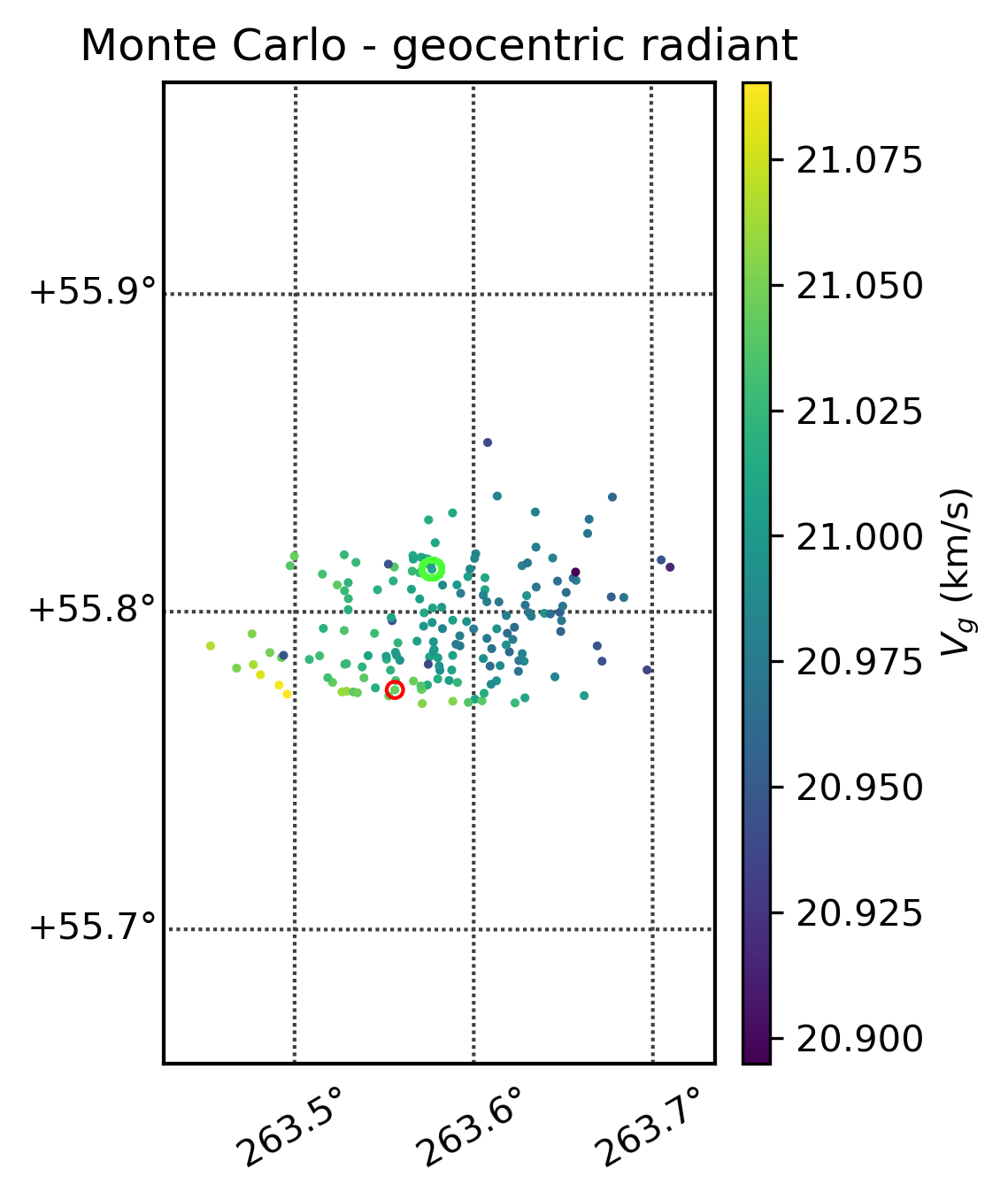}
  \caption{Spread in the geocentric radiant for the modelled Draconid; the geocentric velocity is colour coded. The red circle marks the position of the initial solution ($V_g = $\SI{21.05}{\kilo \metre \per \second}), and the green circle marks the position of the best solution ($V_g = $\SI{21.00}{\kilo \metre \per \second}).}
  \label{fig:mc_vg}
\end{figure}

\begin{figure}
  \includegraphics[width=\linewidth]{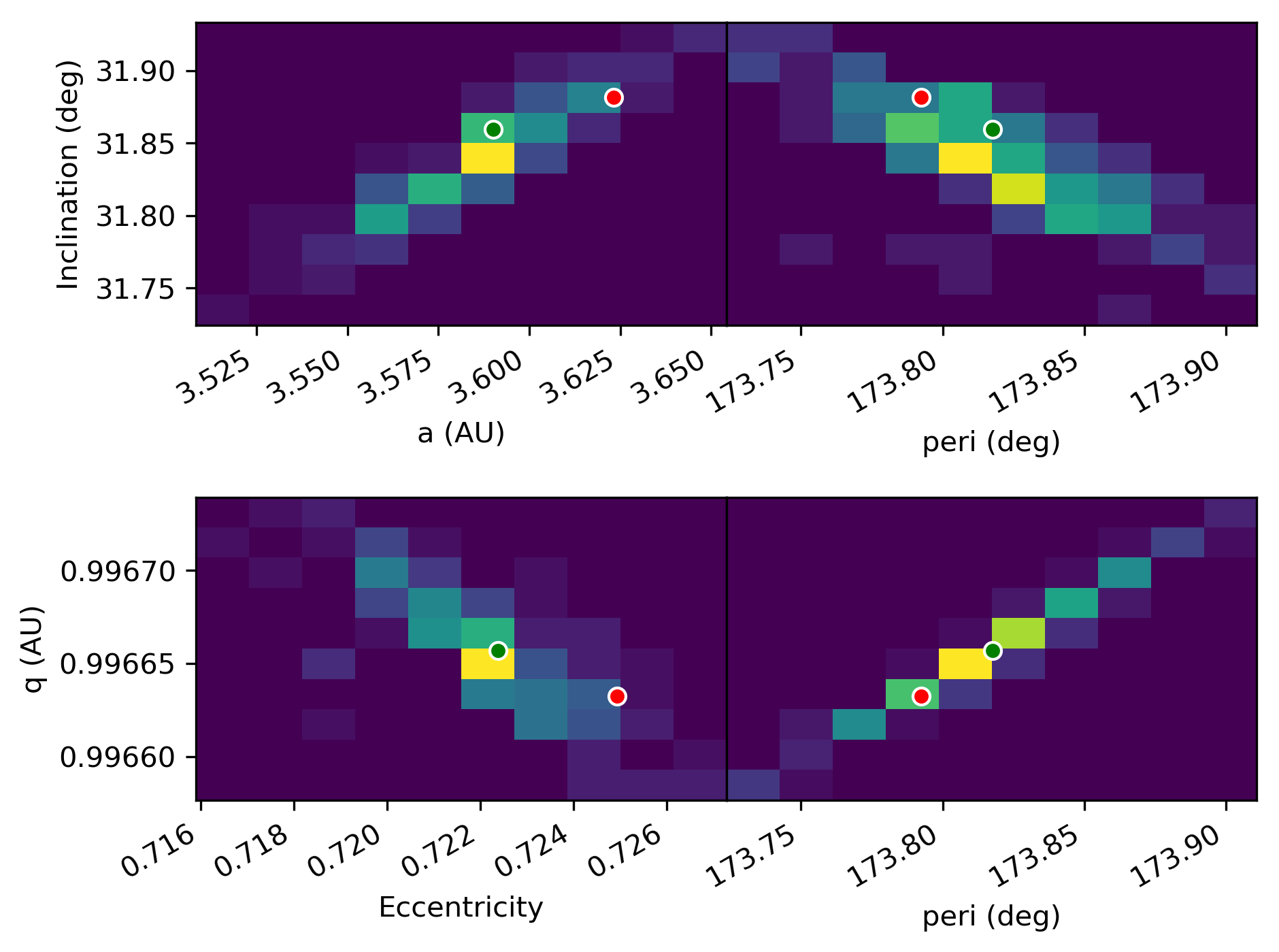}
  \caption{2D histogram of the spread in orbital elements for the modelled Draconid. The red circle marks the position of the initial solution, and the green circle marks the position of the best solution. Brighter bins indicate more trials within the bin.}
  \label{fig:mc_orb_elements}
\end{figure}

Figure \ref{fig:mc_timing_res} shows the geocentric radiants of all Monte Carlo solutions (the value of the square root of the $f_{\Delta t}$ function is color-coded), and figure \ref{fig:mc_vg} shows how the geocentric velocity varies with the radiant position for the example model Draconid meteor. Figure \ref{fig:mc_orb_elements} shows the spread in orbital elements, in particular the strong dependence of individual orbital elements on one another. This behaviour is not captured simply by describing independently computing standard deviations of every orbital element. 

\begin{figure}
  \includegraphics[width=\linewidth]{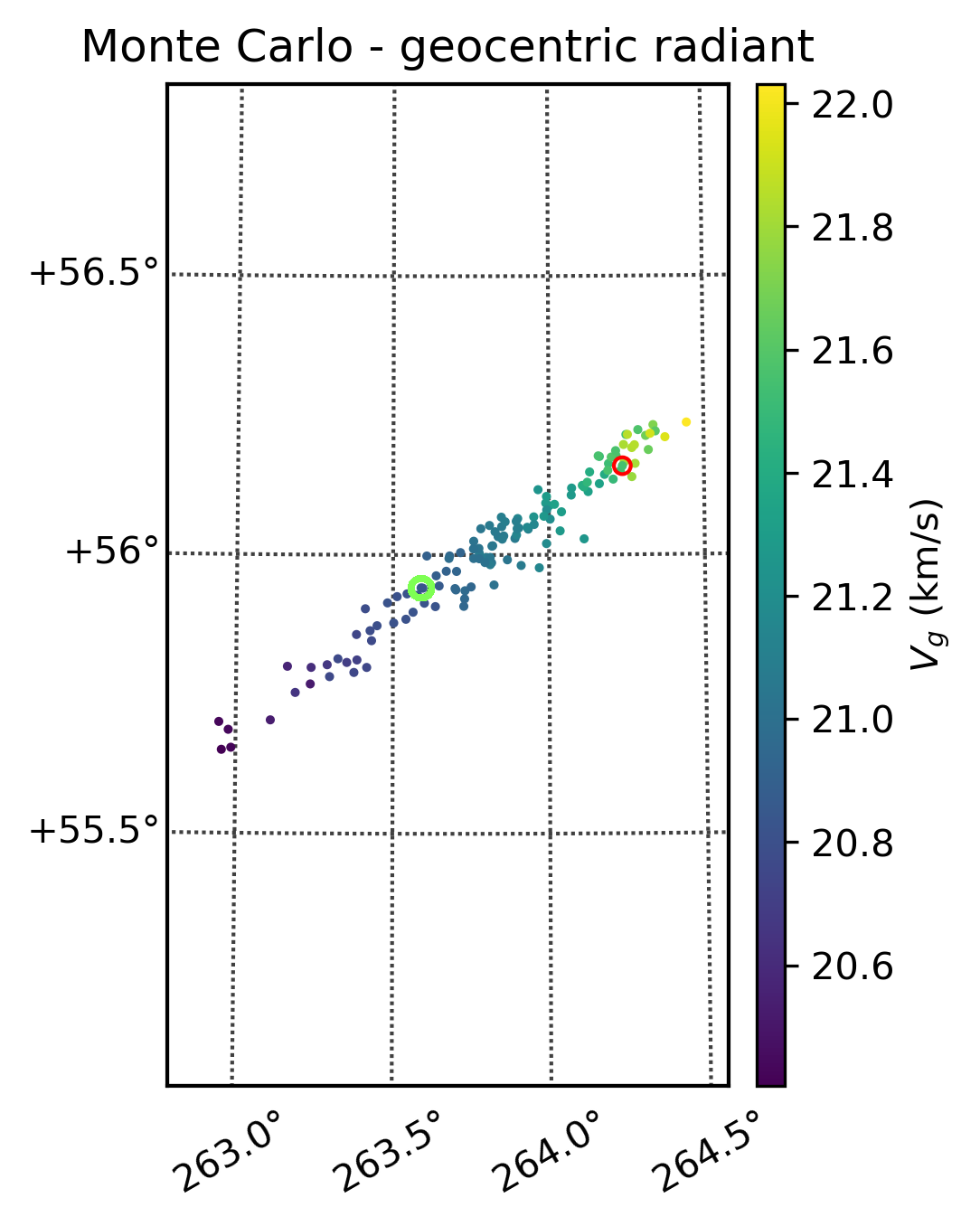}
  \caption{A separate simulation done to illustrate how elongated the radiant uncertainty can be. Here the geocentric velocity is colour coded. The red circle marks the position of the initial (geometrical) solution, and the green circle marks the position of the best (lowest lag cost function) solution. The original model input value of the geocentric velocity was $V_g = \SI{20.893}{\kilo \metre \per \second}$. The initial lines of sight solution underestimated the velocity by $\Delta V_g = \SI{-0.661}{\kilo \metre \per \second}$, while the Monte Carlo method slightly overestimated by only $\Delta V_g = \SI{0.017}{\kilo \metre \per \second}$.}
  \label{fig:mc_vg_elongated}
\end{figure}

To more realistically convey trajectory and orbital uncertainties, we compute covariance matrices of both the orbit and the initial state vector. Note that the uncertainty in the geocentric radiant is not properly represented by considering standard deviations in the right ascension and declination separately. Most two station meteor events, particularly those with a low convergence angle, show an elongated radiant uncertainty. Using a different model Draconid, just such an example is shown in figure \ref{fig:mc_vg_elongated}.

Note that figure \ref{fig:mc_timing_res} shows a clear correlation of the timing residuals (the $f_{\Delta t}$ function) relative to radiant position and a clear global minimum. In experimentation with model fits, we have found this behaviour to be a strong indicator of an improvement in the trajectory solution relative to the geometrical best solution, showing that the best Monte Carlo trajectory should be taken as the solution with lowest lag residuals.

We note that for some model geometries, there are cases when no consistent gradient in the residuals with radiant location is present. In these cases the values of the $f_{\Delta t}$ function are randomly scattered among radiant solutions. In such cases, we found that keeping the original purely geometric solution produced fits closer to the simulated trajectory.

\section{Meteor shower and trajectory simulator}\label{sec:trajsim}

By developing a comprehensive meteor trajectory simulator we wish to generate synthetic measurements for specific video systems in realistic conditions. This involves generating model observations by stipulating real locations of meteor stations, instrument fields of view (FOV), cadence, sensitivity, and measurement uncertainties. In this work we require the simulator to produce simulated trajectories of shower meteors, but sporadic meteors can also be simulated given a sporadic source model. Meteor showers are simulated by specifying the radiant, radiant drift and radiant spread (assumed to be Gaussian), in addition to an activity profile. 

The dynamics of the meteor's motion within the model are generated using the meteor ablation model of \cite{campbell2004model} for which the range of meteoroid masses, the mass index, the meteoroid bulk density distribution and the ablation coefficient are defined as inputs. The attraction of the meteoroid body to the Earth's center due to gravity is taken into account as well. Higher order gravitational coefficients are disregarded because their influence is not measurable using these methods. In what follows, we describe the details of the simulator and demonstrate that it produces meteor trajectories comparable to real observations. The trajectory simulator outputs sets of time, right ascension, declination, and apparent magnitude for every simulated meteor, emulating what would be seen by observers on the ground.

\subsection{Simulating radiants and activity}

For each model station, the following parameters are defined: 
\renewcommand{\labelenumi}{\arabic{enumi}.}
\begin{enumerate}
    \item{The geographical coordinates longitude $\lambda$, geodetic latitude $\varphi$ and elevation above a WGS84 geoid of the Earth $h$}
    \item{The sensor system parameters:}
    \begin{enumerate}
        \item{cadence (i.e. frames per second (FPS) of the video camera)}
        \item{maximum possible deviation in time $\Delta t_{max}$ from the absolute time}
        \item{azimuth $A$ and altitude $a$ of the FOV center for each local site coordinates}
        \item{width and height of the rectangular FOV}
        \item{meteor limiting magnitude $MLM$}
        \item{The radiant power of a zero-magnitude meteor $P_{0m}$ \citep[see][]{ayers1965luminous}}
    \end{enumerate}
\end{enumerate}

For each station, the time offset from the absolute time and asynchronous timing shift between cameras is drawn from a uniform distribution $U(0, \Delta t_{max})$. The time offset and video frame rate are assumed constant over the duration of the meteor. The measurement precision of leading edge picks along the meteor track is simulated by adding Gaussian noise to each simulated measurement with a standard deviation equal to the scatter in residuals for real measurements.

To make the resulting trajectory solution averages per shower have realistic weighted geometries given the station locations, activity profiles for each shower are required. The activity profile of simulated meteor showers is defined by the solar longitude of the peak $\lambda_{\astrosun}^{max}$ and the slope of the activity profile $B$, where the activity is approximated as $ZHR = ZHR_{max} 10^{-B |\lambda_{\astrosun} - \lambda_{\astrosun}^{max}|}$ following \cite{jenniskens1994meteor}. The activity profile is assumed to be symmetric with respect to the peak. $N$ samples are drawn from the activity profile using the inverse sampling transform method - every sample represents one simulated meteor. First, $N$ samples are drawn from a uniform distribution $U(0, 1)$, producing a vector $(y_1, ..., y_N)$. Next, signs are drawn from a uniform distribution $U(-1, 1)$, producing a vector $(s_1, ..., s_N)$. The solar longitude of each sample is then computed as:

\begin{equation}
    \lambda_{\astrosun i} = \lambda_{\astrosun}^{max} + \sign(s_i) \frac{\log_{10} y_i}{B}
\end{equation}

\noindent and only those simulated shower meteors having solar longitudes which occurred between the local astronomical twilight and dawn of all observers are used.

Simulated meteor shower radiants are defined by their geocentric right ascension $\alpha_g$ and declination $\delta_g$ taken to be the mean radiant at the peak together with the standard deviation of the radiant dispersion $\sigma_\alpha$, $\sigma_\delta$. For times away from the peak, the radiant drift $\Delta_\alpha$ and $\Delta_\delta$ in degrees on the sky per degree of solar longitude is used. The shower geocentric velocity $V_g$, and speed dispersion $\sigma_{V_g}$ plus drift $\Delta_{V_g}$ (if known) are also assigned. 

$N$ of individual meteor radiant realizations are drawn from a von Mises distribution (a close approximation to the circular normal distribution) using the centre of distribution at $\mu = 0$ and the dispersion parameter $\kappa = 1/\sigma^2$. $\alpha$ and $\delta$ are drawn independently. This procedure produces vectors $(\alpha'_1, ..., \alpha'_N)$ and $(\delta'_1, ..., \delta'_N)$. These vectors are offsets in right ascension and declination from the mean radiant position. To compute the proper distribution of radiants on the celestial sphere centered around ($\alpha_g$, $\delta_g$), the unit vector $\hat{R_g} = (1, 0, 0)$ is rotated by $-\delta'_i$ on the Y axis, and then by $\alpha'_i$ on the Z axis for every coordinate pair $i$. Next, the resulting vector is rotated by the negative declination of the mean radiant $-\delta_g$ on the Y axis, and then by $\alpha_g$ on the Z axis, and converted to right ascension and declination:

\begin{equation}
\begin{aligned}
    \alpha_{gi} = \atantwo \left( \hat{R}_{giy}, \hat{R}_{gix} \right) \\
    \delta_{gi} = \arcsin \hat{R}_{giz}
\end{aligned}    
\end{equation}

\noindent The radiant drift is applied as:

\begin{equation}
\begin{aligned}
    \alpha_{gi} = \alpha_{gi} + \Delta_\alpha (\lambda_{\astrosun i} - \lambda_{\astrosun}^{max}) \\
    \delta_{gi} = \delta_{gi} + \Delta_\delta (\lambda_{\astrosun i} - \lambda_{\astrosun}^{max})
\end{aligned}    
\end{equation}

Geocentric velocities $V_{gi}$ are drawn from a Gaussian distribution $N(V_g, \sigma_{V_g})$, and a drift in $V_g$ is applied as:

\begin{equation}
    V_{gi} = V_{gi} + \Delta_{V_g} (\lambda_{\astrosun i} - \lambda_{\astrosun}^{max})
\end{equation}

\subsection{Generating meteor state vectors and apparent radiants}

The beginning of the luminous flight of the  meteor is used as the point of reference (i.e. instantaneous measurement of the state vector). This point is randomly generated to be inside the fields of view of at least two stations in the simulation for a given start height. We use a start height of \SI{120}{\kilo \metre} as a reference point between the trajectory and the ablation model. \SI{120}{\kilo \metre} was chosen because almost no meteors end above this height, so the reference point on the trajectory is before or during the luminous phase.

The following paragraphs describe the procedure for generating initial meteor position vectors in 3D space.

Four rays representing the four corners of the FOV of one camera emanate from the coordinates of the station (equivalent to the center of the sensor focal plane). Earth-centered inertial (ECI) coordinates are used. A frustum (truncated pyramid) is obtained by taking 8 points in total, each laying on a FOV corner ray at heights \SI{-5}{\kilo \metre} and \SI[retain-explicit-plus]{+5}{\kilo \metre} around the simulated beginning height for a particular meteor. A random point is generated inside the frustum of one station, and this random sampling is repeated until the point is inside a frustum of at least one another station. The overlap is checked using the quickhull algorithm \citep{barber1996quickhull}. The resulting 3D position vector $\vv{S}$ is taken to be the beginning point of the simulated meteor in ECI coordinates. All initial positions are generated inside overlapping fields of view of at least two cameras due to the computational simplicity of the approach.

From this initial point and the given geocentric radiant, the apparent radiant and the initial velocity is computed in the ECI frame. The initial velocity $v_0$ [\SI{}{\metre \per \second}] is computed from the inverse of the geocentric velocity equation \ref{eq:geocentric_velocity} in Appendix \ref{sec:geocentric_radiant}:

\begin{equation}
    v_0 = \sqrt{v_g^2 + \frac{2*6.67408*5.9722*10^{13}}{|\vv{S}|}}
\end{equation}

\noindent and the apparent values of the radiant $(\alpha_i, \delta_i)$ are numerically inverted using forward mapping equations (see Appendix \ref{sec:geocentric_radiant}). The apparent radiant unit vector $\hat{R}$ is computed by converting the spherical coordinates $(\alpha_i, \delta_i)$ to their ECI components using equation \ref{eq:radec2cartesian}. Note that the $v_g$ is converted into the initial velocity by assuming that the stations are moving in the ECI coordinates, and thus the whole coordinate system rotates with the Earth, making a correction to the meteor velocity for Earth's rotation unnecessary; such a correction would be needed for an ECEF treatment. Radiants with zenith angles $z_c > \ang{80}$ are skipped to avoid simulating meteors which do not propagate down in the atmosphere.

\subsection{Simulating meteoroid dynamics} \label{subsec:dynamics_sim}

To simulate realistic meteor dynamics, the meteoroid ablation model of \cite{campbell2004model} is used. For each shower, a range of visible masses $m_{min}$, $m_{max}$ and a mass index $s$ based on literature values for a particular shower are defined. The masses are sampled using inverse transform sampling from the cumulative number as a function of mass distribution:

\begin{equation}
    f(m) = m^{1 - s}
\end{equation}

Meteoroid densities are either sampled uniformly from a user defined range, or using density distributions given by \cite{moorhead2017two}. The apparent ablation coefficient $\sigma$  (usually given in [\SI{}{\square \second \per \square \kilo \metre}]) is applied in the ablation model through modification of the energy needed to ablate a unit mass $L$ [\SI{}{\joule \per \kilogram}], (make sure to convert $\sigma$ to \SI{}{\square \second \per \square \metre}) which is computed as:

\begin{equation} \label{eq:heat_of_ablation}
    L = \frac{\Lambda}{2 \sigma \Gamma}
\end{equation}

\noindent where $\Lambda = 0.5$ is the heat transfer coefficient, and $\Gamma = 1.0$ is the drag coefficient. Note that in the field of aerodynamics the notation $C_d$ is used for the drag coefficient, where $\Gamma = 2 C_d$. The ablation model provides vectors of height, length, and luminosity along the meteor path from the beginning point with a temporal resolution of \SI{0.001}{\second}. Note that $\sigma$ is used throughout the text with different meanings. In equation \ref{eq:heat_of_ablation} it is used for the ablation coefficient, while at all other places it is used for standard deviation.

\subsection{Generating synthetic trajectory data}

The duration $t_{meteor}$ of a meteor is obtained from the ablation model. We assume the beginning time $t = 0$ corresponds to a given solar longitude for the corresponding reference Julian date $JD_{ref}$. A vector of times is obtained by sampling the range $(0, t_{meteor})$ with the step $1/FPS$.

The instantaneous model luminosity $I$ at a given time is converted to a range-corrected apparent magnitude $M_v$ and only those points above the meteor limiting magnitude of individual stations are taken:

\begin{equation}
\begin{aligned}
    M_A = -2.5 \log_{10} \frac{I}{P_{0m}} \\
    M_v = M_A - 5 \log_{10} \frac{10^5}{r}
\end{aligned}    
\end{equation}

\noindent where $M_A$ is the absolute magnitude (magnitude at \SI{100}{\kilo \metre} range) and $r$ is the range in meters from the station to the meteor. $P_{0m}$ is the power of a zero magnitude meteor for the appropriate bandpass taken from \cite{weryk2013simultaneous}. No correction for angular velocity or extinction loss is included.

The 3D meteor positions are projected to local spherical coordinates of stations to generate synthetic observations. We simulate the real movement of the stations due to Earth's rotation by computing ECI coordinates $\vv{ECI_j}$ of stations at every model point in time $t_k$. The position of the meteor in ECI coordinates at time $t_j$ is computed as:

\begin{equation}
    \vv{T_j} = \vv{S} - d(t_j) \hat{R}
\end{equation}

\noindent where $\vv{S}$ is the initial position at $t = 0$, and $\hat{R}$ is the apparent radiant unit vector in ECI coordinates. The additional decrease in height due to Earth's gravity is applied using equation \ref{eq:tkj}, where $\Delta h (t_j)$ is the decrease in height at every point in time due to gravity since the beginning (in meters). This procedure simulates the curvature of the trajectory due to gravity, assuming the pull is perpendicular to the WGS84 reference ellipsoid. $\Delta h (t_j)$ is computed as described in Appendix \ref{appendix:gravity_drop}. A unit vector pointing from the station to the position of the meteor on the trajectory is computed as:

\begin{equation}
    \hat{r} = \frac{ \vv{T_j} - \vv{ECI_j} }{| \vv{T_j} - \vv{ECI_j} |}
\end{equation}

We simulate the observational precision of a system by adding Gaussian noise with a standard deviation $\sigma$, derived from real measurements of the actual systems, to the synthetic observations. We separate the vector $\hat{r}$ into orthogonal components $\hat{u}$ and $\hat{v}$:

\begin{equation}
\begin{aligned}
    \hat{z} = [0, 0, 1]\\
    \hat{u} = \frac{ \hat{r} \times \hat{z} }{|\hat{r} \times \hat{z}|} \\
    \hat{v} = \frac{ \hat{u} \times \hat{r} }{|\hat{u} \times \hat{r}|}
\end{aligned}    
\end{equation}

\noindent The direction vector (all in ECI) with the added noise is then:

\begin{equation} \label{eq:adding_gaussian_noise}
    \vv{r'} = \hat{r} + \mathcal{N} \left(0, \sigma \right) \hat{u} + \mathcal{N} \left(0, \sigma \right) \hat{v}
\end{equation}

\noindent where $\mathcal{N} \left(0, \sigma \right)$ is a scalar drawn from a Gaussian distribution with a mean of 0 and a standard deviation of $\sigma$. The samples are drawn separately for each term. The direction vector is converted to equatorial coordinates in the epoch of date:

\begin{equation}
\begin{aligned}
    \hat{r'} = \frac{ \vv{r'} }{ |\vv{r'}| } \\
    \alpha_j = \atantwo \left ( \hat{r'}_y, \hat{r'}_x \right) \\
    \delta_j = \arcsin{\hat{r'}_z}
\end{aligned}    
\end{equation}

\noindent Finally, the appropriate timing offset $\Delta t$ for a given station (randomized on a per meteor basis) is added to time $t_j$, completing the set of synthetic measurements for each simulated meteor. 

At the end of this procedure one obtains a set of $N_{meas}$ synthetic measurements from every station for every generated meteor. Synthetic meteors are uniquely defined by the Julian date of their beginning $JD_{ref}$, set of relative times since the beginning $(t_0, ..., t_j)$, a set of right ascensions $(\alpha_0, ..., \alpha_j)$ and declinations $(\delta_0, ..., \delta_j)$ in the epoch of date. Note that the epoch here is not J2000; to avoid confusion we convert the model measurements to local azimuth $(A_0, ..., A_j)$ and altitude $(a_0, ..., a_j)$ in the epoch of date from a particular station using equations given in Appendix \ref{appendix:altaz2radec}.

Although the simulator reproduces many features of the observed data, a major difference with real meteors is that synthetic trajectories all start within the FOVs of at least 2 stations. It is not clear that this limitation is significant for the current work. While this might be alleviated by generating the state vectors slightly outside the FOV of one camera this would be at the expense of having to compute the propagation as well, which would significantly increase the computational load of finding a synthetic meteor that is actually visible from 2 or more stations.

\section{Conclusion}

A novel Monte Carlo meteor trajectory method was developed which takes the dynamics of meteors into account without assuming any formulated meteor propagation model. This leverages the fact that modern meteor electro-optical systems have sufficient precision to routinely record deceleration, allowing an entirely independent check on the solution consistency between stations. 

Improvements in weighting multi-station observations as well as a new method of initial velocity estimation have been proposed. A limitation of the new Monte Carlo solver is that it does not work for meteors with no temporal overlap between stations. In those cases a dynamical model must be used to estimate timing differences and the velocity, but the radiant and its uncertainty can be estimated using purely geometrical methods, similar to earlier approaches \citep{weryk2012simultaneous, gural2012solver}.

We develop a meteor trajectory simulator which uses a numerical meteor ablation model to simulate meteor dynamics. The simulator will be used in the second paper in this series to investigate radiant and velocity accuracy that can be achieved for various real-world optical systems and meteor showers.

Finally, we provide a detailed set of equations and explanations for estimating meteor trajectories and computing orbits starting just from a set of multi-station observations. We also have made the associated code-base openly available for all to use. Additional details are included in the accompanying appendices. An improved version of the MPF method incorporating the findings of this paper will be published in the future. We invite readers to continue to the second paper in this series for results.

\subsection{Note on code availability}

Implementation of the meteor simulator as well as implementation of all meteor solvers used in this work are published as open source on the following GitHub web page: \url{https://github.com/wmpg/WesternMeteorPyLib}. Readers are encouraged to contact the authors in the event they are not able to obtain the code on-line.

\section{Acknowledgements}

We thank Dr. Eleanor Sansom for a helpful and detailed review of an earlier version of this manuscript. Also, we thank Dr. Steven Ehlert for suggesting the name "PyLIG" for the Monte Carlo trajectory solver developed as a part of this work, as well as David L. Clark for a fruitful discussion about meteor orbit computation. Funding for this work was provided by the NASA Meteoroid Environment Office under cooperative agreement 80NSSC18M0046. PGB acknowledges funding support from the Canada Research Chair program and the Natural Sciences and Engineering Research Council of Canada.




\bibliographystyle{mnras}
\bibliography{bibliography} 

\begin{thebibliography}{}
\makeatletter
\relax
\def\mn@urlcharsother{\let\do\@makeother \do\$\do\&\do\#\do\^\do\_\do\%\do\~}
\def\mn@doi{\begingroup\mn@urlcharsother \@ifnextchar [ {\mn@doi@}
  {\mn@doi@[]}}
\def\mn@doi@[#1]#2{\def\@tempa{#1}\ifx\@tempa\@empty \href
  {http://dx.doi.org/#2} {doi:#2}\else \href {http://dx.doi.org/#2} {#1}\fi
  \endgroup}
\def\mn@eprint#1#2{\mn@eprint@#1:#2::\@nil}
\def\mn@eprint@arXiv#1{\href {http://arxiv.org/abs/#1} {{\tt arXiv:#1}}}
\def\mn@eprint@dblp#1{\href {http://dblp.uni-trier.de/rec/bibtex/#1.xml}
  {dblp:#1}}
\def\mn@eprint@#1:#2:#3:#4\@nil{\def\@tempa {#1}\def\@tempb {#2}\def\@tempc
  {#3}\ifx \@tempc \@empty \let \@tempc \@tempb \let \@tempb \@tempa \fi \ifx
  \@tempb \@empty \def\@tempb {arXiv}\fi \@ifundefined
  {mn@eprint@\@tempb}{\@tempb:\@tempc}{\expandafter \expandafter \csname
  mn@eprint@\@tempb\endcsname \expandafter{\@tempc}}}

\bibitem[\protect\citeauthoryear{Ayers}{Ayers}{1965}]{ayers1965luminous}
Ayers W.~G.,  1965

\bibitem[\protect\citeauthoryear{Barber, Dobkin  \& Huhdanpaa}{Barber
  et~al.}{1996}]{barber1996quickhull}
Barber C.~B.,  Dobkin D.~P.,   Huhdanpaa H.,  1996, ACM Transactions on
  Mathematical Software (TOMS), 22, 469

\bibitem[\protect\citeauthoryear{Borovi{\v{c}}ka}{Borovi{\v{c}}ka}{1990}]{borovicka1990comparison}
Borovi{\v{c}}ka J.,  1990, Bulletin of the Astronomical Institutes of
  Czechoslovakia, 41, 391

\bibitem[\protect\citeauthoryear{Borovi{\v{c}}ka, Spurn{\`y}  \&
  Koten}{Borovi{\v{c}}ka et~al.}{2007}]{borovivcka2007atmospheric}
Borovi{\v{c}}ka J.,  Spurn{\`y} P.,   Koten P.,  2007, Astronomy \&
  Astrophysics, 473, 661

\bibitem[\protect\citeauthoryear{Burke}{Burke}{1986}]{Burke1986}
Burke J.~G.,  1986, Cosmic debris: Meteorites in history.
Univ of California Press

\bibitem[\protect\citeauthoryear{Campbell-Brown \& Koschny}{Campbell-Brown \&
  Koschny}{2004}]{campbell2004model}
Campbell-Brown M.,  Koschny D.,  2004, Astronomy \& Astrophysics, 418, 751

\bibitem[\protect\citeauthoryear{Ceplecha}{Ceplecha}{1979}]{Ceplecha1979a}
Ceplecha Z.,  1979, Bulletin of the Astronomical Institutes of Czechoslovakia,
  30, 349

\bibitem[\protect\citeauthoryear{Ceplecha}{Ceplecha}{1987}]{ceplecha1987geometric}
Ceplecha Z.,  1987, Bulletin of the Astronomical Institutes of Czechoslovakia,
  38, 222

\bibitem[\protect\citeauthoryear{Ceplecha, Borovi{\v{c}}ka, Elford, ReVelle,
  Hawkes, Porub{\v{c}}an  \& {\v{S}}imek}{Ceplecha
  et~al.}{1998}]{ceplecha1998meteor}
Ceplecha Z.,  Borovi{\v{c}}ka J.,  Elford W.~G.,  ReVelle D.~O.,  Hawkes R.~L.,
   Porub{\v{c}}an V.,   {\v{S}}imek M.,  1998, Space Science Reviews, 84, 327

\bibitem[\protect\citeauthoryear{Clark}{Clark}{2010}]{clark2010searching}
Clark D.~L.,  2010, PhD thesis, The University of Western Ontario, London,
  Ontario, Canada

\bibitem[\protect\citeauthoryear{Eberhart \& Kennedy}{Eberhart \&
  Kennedy}{1995}]{eberhart1995new}
Eberhart R.,  Kennedy J.,  1995, in Micro Machine and Human Science, 1995.
  MHS'95., Proceedings of the Sixth International Symposium on. pp 39--43

\bibitem[\protect\citeauthoryear{Eberly}{Eberly}{2006}]{eberly20063d}
Eberly D.~H.,  2006, 3D game engine design: a practical approach to real-time
  computer graphics.
CRC Press

\bibitem[\protect\citeauthoryear{Egal, Gural, Vaubaillon, Colas  \&
  Thuillot}{Egal et~al.}{2017}]{egal2017challenge}
Egal A.,  Gural P.,  Vaubaillon J.,  Colas F.,   Thuillot W.,  2017, Icarus

\bibitem[\protect\citeauthoryear{Folkner, Williams, Boggs, Park  \&
  Kuchynka}{Folkner et~al.}{2014}]{folkner2014planetary}
Folkner W.~M.,  Williams J.~G.,  Boggs D.~H.,  Park R.~S.,   Kuchynka P.,
  2014, Interplanet. Netw. Prog. Rep 196, C1

\bibitem[\protect\citeauthoryear{Gural}{Gural}{2001}]{gural2001fully}
Gural P.,  2001, WGN, Journal of the International Meteor Organization, 29, 134

\bibitem[\protect\citeauthoryear{Gural}{Gural}{2012}]{gural2012solver}
Gural P.~S.,  2012, Meteoritics \& Planetary Science, 47, 1405

\bibitem[\protect\citeauthoryear{Hawkes \& Jones}{Hawkes \&
  Jones}{1975}]{Hawkes1975}
Hawkes R.,  Jones J.,  1975, Monthly notices of the Royal Astronomical Society,
  173, 339

\bibitem[\protect\citeauthoryear{Hughes}{Hughes}{1982}]{hughes1982history}
Hughes D.~W.,  1982, Vistas in astronomy, 26, 325

\bibitem[\protect\citeauthoryear{Jenniskens}{Jenniskens}{1994}]{jenniskens1994meteor}
Jenniskens P.,  1994, Astronomy and Astrophysics, 287, 990

\bibitem[\protect\citeauthoryear{Jenniskens, Gural, Dynneson, Grigsby, Newman,
  Borden, Koop  \& Holman}{Jenniskens et~al.}{2011}]{jenniskens2011cams}
Jenniskens P.,  Gural P.,  Dynneson L.,  Grigsby B.,  Newman K.,  Borden M.,
  Koop M.,   Holman D.,  2011, \mn@doi [Icarus]
  {https://doi.org/10.1016/j.icarus.2011.08.012}, 216, 40

\bibitem[\protect\citeauthoryear{Meeus}{Meeus}{1998}]{meeus1998astronomical}
Meeus J.~H.,  1998, Astronomical algorithms, 2 edn.
Willmann-Bell, Incorporated

\bibitem[\protect\citeauthoryear{Moorhead, Blaauw, Moser, Campbell-Brown, Brown
   \& Cooke}{Moorhead et~al.}{2017}]{moorhead2017two}
Moorhead A.~V.,  Blaauw R.~C.,  Moser D.~E.,  Campbell-Brown M.~D.,  Brown
  P.~G.,   Cooke W.~J.,  2017, Monthly Notices of the Royal Astronomical
  Society, 472, 3833

\bibitem[\protect\citeauthoryear{Pavlis, Holmes, Kenyon  \& Factor}{Pavlis
  et~al.}{2012}]{pavlis2012development}
Pavlis N.~K.,  Holmes S.~A.,  Kenyon S.~C.,   Factor J.~K.,  2012, Journal of
  geophysical research: solid earth, 117

\bibitem[\protect\citeauthoryear{Romig}{Romig}{1966}]{romig1966scientific}
Romig M.~F.,  1966, Meteoritics \& Planetary Science, 3, 11

\bibitem[\protect\citeauthoryear{Sansom, Rutten  \& Bland}{Sansom
  et~al.}{2017}]{sansom2017analyzing}
Sansom E.~K.,  Rutten M.~G.,   Bland P.~A.,  2017, The Astronomical Journal,
  153, 87

\bibitem[\protect\citeauthoryear{Schiaparelli \& von Boguslawski}{Schiaparelli
  \& von Boguslawski}{1871}]{schiaparelli1871entwurf}
Schiaparelli G.~V.,  von Boguslawski G.,  1871, Entwurf einer astronomischen
  Theorie der Sternschnuppen.
Th. von der Nahmer

\bibitem[\protect\citeauthoryear{Stokan, Campbell-Brown, Brown, Hawkes, Doubova
   \& Weryk}{Stokan et~al.}{2013}]{stokan2013}
Stokan E.,  Campbell-Brown M.~D.,  Brown P.,  Hawkes R.,  Doubova M.,   Weryk
  R.,  2013, \mn@doi [Monthly Notices of the Royal Astronomical Society]
  {10.1093/mnras/stt779}, 433, 962

\bibitem[\protect\citeauthoryear{Subasinghe, Campbell-Brown  \&
  Stokan}{Subasinghe et~al.}{2016}]{subasinghe2016physical}
Subasinghe D.,  Campbell-Brown M.~D.,   Stokan E.,  2016, \mn@doi [Monthly
  Notices of the Royal Astronomical Society] {10.1093/mnras/stw019}, 457, 1289

\bibitem[\protect\citeauthoryear{Subasinghe, Campbell-Brown  \&
  Stokan}{Subasinghe et~al.}{2017}]{subasinghe2017luminous}
Subasinghe D.,  Campbell-Brown M.,   Stokan E.,  2017, \mn@doi [Planetary and
  Space Science] {https://doi.org/10.1016/j.pss.2016.12.009}, pp~--

\bibitem[\protect\citeauthoryear{Tsuchiya, Sato, Watanabe, Moorhead, Moser,
  Brown  \& Cooke}{Tsuchiya et~al.}{2017}]{tsuchiya2017correction}
Tsuchiya C.,  Sato M.,  Watanabe J.-i.,  Moorhead A.~V.,  Moser D.~E.,  Brown
  P.~G.,   Cooke W.~J.,  2017, Planetary and Space Science, 143, 142

\bibitem[\protect\citeauthoryear{Vida, Brown, Campbell-Brown  \& Huggins}{Vida
  et~al.}{2018a}]{vida2018canadian}
Vida D.,  Brown P.,  Campbell-Brown M.,   Huggins S.,  2018a, in International
  Meteor Conference, Petnica, Serbia, 21-24 September 2017. pp 18--24

\bibitem[\protect\citeauthoryear{Vida, Brown  \& Campbell-Brown}{Vida
  et~al.}{2018b}]{vida2018modeling}
Vida D.,  Brown P.~G.,   Campbell-Brown M.,  2018b, Monthly Notices of the
  Royal Astronomical Society, 479, 4307

\bibitem[\protect\citeauthoryear{Weryk \& Brown}{Weryk \&
  Brown}{2012}]{weryk2012simultaneous}
Weryk R.~J.,  Brown P.~G.,  2012, \mn@doi [Planetary and Space Science]
  {https://doi.org/10.1016/j.pss.2011.12.023}, 62, 132

\bibitem[\protect\citeauthoryear{Weryk \& Brown}{Weryk \&
  Brown}{2013}]{weryk2013simultaneous}
Weryk R.~J.,  Brown P.~G.,  2013, \mn@doi [Planetary and Space Science]
  {https://doi.org/10.1016/j.pss.2013.03.012}, 81, 32

\bibitem[\protect\citeauthoryear{Weryk, Campbell-Brown, Wiegert, Brown,
  Krzeminski  \& Musci}{Weryk et~al.}{2013}]{weryk2013camo}
Weryk R.,  Campbell-Brown M.,  Wiegert P.,  Brown P.,  Krzeminski Z.,   Musci
  R.,  2013, \mn@doi [Icarus] {https://doi.org/10.1016/j.icarus.2013.04.025},
  225, 614

\bibitem[\protect\citeauthoryear{Whipple \& Jacchia}{Whipple \&
  Jacchia}{1957}]{whipple1957reduction}
Whipple F.~L.,  Jacchia L.~G.,  1957, Smithsonian Contributions to
  Astrophysics, 1, 183

\makeatother
\end{thebibliography}


\appendix

\section{Bending of the trajectory due to gravity} \label{appendix:gravity_drop}

The straight line approximation for trajectories breaks down in the case of long (> \SI{4}{\second}) meteors, when they will show vertical curvature that should be visible even with less precise systems. At a height of \SI{100}{\kilo \metre} the gravitational acceleration is $g = \SI{\sim 9.5}{\metre \per \square \second}$, although it changes as the meteor descends through the atmosphere with the classical relation:

\begin{equation} \label{eq:gravity_acceleration_classical}
    g(r) = \frac{G M_E}{r^2}
\end{equation}

\noindent where $M_E$ is the mass of the Earth and $r$ is the distance of the meteor from the centre of the Earth. To compute the changing value of the gravitational acceleration, we assume that at the begin point, the downward vertical component of the meteor's velocity $v_z$ is equal to the vertical component of the initial velocity:

\begin{equation}
    v_z = -v_0 \cos{z_c}
\end{equation}

\noindent where $v_0$ is the initial velocity and $z_c$ the apparent zenith angle. Thus, the gravitational acceleration at a relative time $t$ after the beginning of the meteor is:

\begin{equation}
    g(t) = \frac{G M_E}{(r_0 + v_z t)^2}
\end{equation}

\noindent where $M_E$ is the mass of the Earth and $r_0$ is the distance from the centre of the Earth to the beginning height of the meteor. The total drop of the meteor due to gravity after time $T$ is then:

\begin{equation}
    \Delta h(T) = \int_0^T g(t) t \, dt
\end{equation}

\noindent After integration we obtain the following relation:

\begin{equation}
    \Delta h (T) = \frac{G M_E}{v_z^2} \left ( \frac{r_0}{r_0 + v_z T} + \ln{\frac{r_0 + v_z T}{r_0}} - 1\right )
\end{equation}

To avoid domain issues when $v_z \approx 0$ we only use this expanded equation if $|v_z| > \SI{100}{\metre \per \second}$, otherwise we use equation \ref{eq:gravity_acceleration_classical} with $r = r_0$ to compute $g$ and the classical way of computing the additional drop in height due to gravity:

\begin{equation}
    \Delta h (T) = \frac{1}{2} g T^2
\end{equation}

Applying $\Delta h$ to the vertical component of the meteor at every point in time effectively simulates the curvature of the meteor's trajectory due to gravity.

\section{Distance between lines in 3D space} \label{appendix:3d_line_dist}

Let vector $\vv{P}$ be the position of the observer in an arbitrary rectangular coordinate system, and $\vv{U}$ be the direction vector of the line of sight emanating from the observer. Let $\vv{S}$ be the position of the state vector, and $\vv{R}$ be the radiant vector. The closest points of approach can be calculated as:
    
    \begin{equation}
    \begin{aligned}
        \vv{w} = \vv{P} - \vv{S} \\
        a = \vv{U} \cdot \vv{U} \\
        b = \vv{U} \cdot \vv{R} \\
        c = \vv{R} \cdot \vv{R} \\ 
        d = \vv{U} \cdot \vv{w} \\
        e = \vv{R} \cdot \vv{w} \\
        Q_C = \frac{b e - c d}{a c - b^2} \\
        T_C = \frac{a e - b d}{a c - b^2} \\
        \vv{Q} = \vv{P} + Q_C \vv{U} \\
        \vv{T} = \vv{S} + T_C \vv{R} \\
        d = | \vv{Q} - \vv{T} | \\
    \end{aligned}
    \end{equation}
    
\noindent where $\vv{Q}$ is the point on the observer's line of sight closest to the radiant line, and $\vv{T}$ is the point on the radiant line closest to the line of sight of the observer. $d$ is the distance between those two points. The equations are taken from \cite{eberly20063d} in a modified form.

\section{Orbit computation} \label{appendix:orbit}

The orbit is computed from 4 parameters: The apparent radiant unit vector $\hat{R}$, the initial velocity $v_{0}$, the ECI coordinates of the state vector $\vv{S}$, and the reference Julian date of the beginning of the meteor $JD_{ref}$. The equations below assume that the radiant and the state vector are given in the epoch of date, not J2000. Furthermore, we assume that the location of the state vector is at the beginning of the meteor, not at an average point on the trajectory. The state vector $\vv{S}$ should be in meters and the initial velocity $v_{0}$ in \SI{}{\metre \per \second} to be consistent with constants and parameter units used herein.
    
First, the geocentric latitude of the state vector is calculated as:

\begin{equation}
    \varphi' = \atantwo \left ( S_z, \sqrt{S_x^2 + S_y^2} \right)
\end{equation}

Next, care must taken to use the Barycentric Dynamical Time $TDB$ in calculations where necessary. For epochs in 1972 and later the dynamical time is simply calculated as the Julian date with the added leap seconds $\Delta t$ up to the given $JD$, plus a constant of \SI{32.184}{\second} \citep{clark2010searching}. The number of leap seconds can be obtained from the United States Naval observatory FTP site\footnote{USNO leap seconds file, \url{ftp://maia.usno.navy.mil/ser7/tai-utc.dat}, (Accessed February 18, 2018)}. For example, $\Delta t$ for a meteor observed between 2006 and 2009 is \SI{33}{\second}, while for a meteor observed after January 1, 2017  (until a future leap second is added) is \SI{37}{\second}.

\begin{equation}
    TDB = JD_{ref} + \frac{\Delta t + 32.184}{86400}
\end{equation}

Next, the geodetic latitude $\varphi$ and the longitude $\lambda$ of the beginning point of the meteor projected onto the Earth's surface are calculated from the ECI coordinates of the state vector using the method described in \ref{appendix:eci2geo}.

\subsection{Correcting the apparent radiant and the velocity for Earth's rotation} \label{subsec:rotation_correction}

If the trajectory was estimated with the intersecting planes method, or if the stations were kept fixed, one needs to correct the radiant for Earth's rotation. Please note the important fact that the correction described in this section must not be applied if the ECI coordinates of the stations were moving in time in the trajectory estimation procedure. Thus, if the station coordinates were moving during the meteor event, the velocity vector is simply calculated as:

\begin{equation}
    \vv{v_0} = v_0 \hat{R}
\end{equation}

\noindent and the rest of the equations in this subsection \ref{subsec:rotation_correction} can be skipped. Otherwise, the procedure described below must be followed.

The rotation velocity of the Earth (in \SI{}{\metre \per \second}) at the height of the state vector can be calculated as:

\begin{equation}
    v_e = \frac{2 \pi |\vv{S}| \cos \varphi'}{86164.09053} 
\end{equation}

\noindent where the number in the denominator is the duration of the sidereal day in seconds.

Next, as the direction of the Earth's rotation vector is always towards the east, we can calculate the components of the velocity vector of the meteor $\vv{v_0}$ as:

\begin{equation}
\begin{aligned}
    v_{0x} = v_0 \hat{R}_x - v_e \cos \alpha_e \\
    v_{0y} = v_0 \hat{R}_y - v_e \sin \alpha_e \\
    v_{0z} = v_0 \hat{R}_z
\end{aligned}
\end{equation}

\noindent where $\alpha_e$ is the right ascension of the direction of the rotation of the Earth. This can be calculated using the equations given in Appendix \ref{appendix:altaz2radec} if we take the azimuth to be $A = \frac{\pi}{2}$  (i.e. due East) and elevation $a = 0$.

It is very important to note that this correction only influences the direction of the radiant, but not the initial velocity itself. This is only true if ECI coordinates are used throughout, regardless of keeping the stations fixed or not.

\subsection{Geocentric radiant} \label{sec:geocentric_radiant}

First, we calculate equatorial coordinates of the apparent radiant following \cite{ceplecha1987geometric}:

\begin{equation} \label{eq:eci2radec}
\begin{aligned}
    \hat{v_0} = \frac{\vv{v_0}}{|\vv{v_0}|} \\
    \alpha = \atantwo \left ( \hat{v}_{0y}, \hat{v}_{0x} \right ) \\
    \delta = \arcsin \hat{v}_{0z}
\end{aligned}
\end{equation}

\noindent The geocentric velocity is calculated as:

\begin{equation} \label{eq:geocentric_velocity}
    v_g = \sqrt{v_0^2 - \frac{2*6.67408*5.9722*10^{13}}{|\vv{S}|}}
\end{equation}

\noindent where the second term under the square root is the square of the escape velocity $\left ( \frac{2 G M_E}{r} \right )$ at the height of the state vector. Next, the zenith attraction correction is applied using the Schiaparelli method \citep{gural2001fully}:

\begin{equation}
\begin{aligned}
    z_c = \arccos \left ( \sin \delta \sin \varphi' + \cos \delta \cos \varphi' \cos \left ( \theta' - \alpha \right ) \right ) \\
    \Delta z_c = 2 \atantwo \left ( (v_0 - v_g) \tan \frac{z_c}{2}, v_0 + v_g \right) \\
    z_g = z_c + |\Delta z_c|
\end{aligned}
\end{equation}

\noindent where $z_c$ is the apparent zenith angle, $\theta'$ is the apparent local sidereal time (see Appendix \ref{appendix:lst}), $\Delta z_c$ the zenith attraction correction, and $z_g$ the zenith angle of the geocentric radiant.

The azimuth $A_c$ of the radiant (possibly corrected for Earth's rotation) is calculated using the equations given in Appendix \ref{appendix:radec2altaz}. The apparent $\alpha$ and $\delta$ should be used, and care must be taken to use the geocentric latitude $\varphi'$ instead of the geodetic latitude. The geocentric radiant in equatorial coordinates ($\alpha_g$, $\delta_g$) is then calculated using the equations given in Appendix \ref{appendix:altaz2radec}, where the azimuth is $A = A_c$, the elevation is $a = \frac{\pi}{2} - z_g$, and the geocentric latitude $\varphi'$ must be used as well.

Next, the radiant is precessed from the epoch of date ($JD_{ref}$) to J2000 using the equations given in Appendix \ref{appendix:precession_eq}. The geocentric ecliptic longitude $\lambda_g$ and latitude $\beta_g$ are calculated with equations given in Appendix \ref{appendix:ecliptic}; care must be taken to use the Julian date of J2000 ($JD = 2451545$) when computing ecliptic coordinates, not $JD_{ref}$.

\subsection{Precessing ECI coordinates to J2000}

As the ECI coordinates of the meteor are in the epoch of date, they have to be precessed to J2000. This can easily be done by converting them to spherical coordinates:

\begin{equation}
\begin{aligned}
    r_{ECI} = |\vv{S}| \\
    \alpha_{ECI} = \atantwo \left ( S_y, S_x \right) \\
    \delta_{ECI} = \arccos \frac{S_z}{r_{ECI}}
\end{aligned}
\end{equation}

\noindent were $r_{ECI}$ is the distance from the center of the Earth to the reference position of the meteor, and $\alpha_{ECI}$ and $\delta_{ECI}$ are angular components. $\alpha_{ECI}$ and $\delta_{ECI}$ are precessed to J2000 from $JD_{ref}$ using equations given in Appendix \ref{appendix:precession_eq}, after which $\alpha'_{ECI}$ and $\delta'_{ECI}$ are obtained. Finally, these coordinates can be converted back to rectangular ECI coordinates in J2000:

\begin{equation}
\begin{aligned}
    S'_x = r_{ECI} \sin \delta'_{ECI} \cos \alpha'_{ECI} \\
    S'_y = r_{ECI} \sin \delta'_{ECI} \sin \alpha'_{ECI} \\
    S'_z = r_{ECI} \cos \delta'_{ECI}
\end{aligned}
\end{equation}

\subsection{Position and the velocity of the Earth}

JPL DE430 ephemerids \citep{folkner2014planetary} are used for computing Cartesian heliocentric ecliptic coordinates and the velocity of the Earth at the reference dynamical time $TDB$. As the implementation of the ephemerids does not allow the calculation of the heliocentric ecliptic coordinates of the Earth directly, the following procedure was adopted:

\begin{enumerate}
    \item The position $\vv{R_{EMB}}$ and the velocity $\vv{V_{EMB}}$ of the Earth-Moon barycentre with respect to the Solar System Barycentre is obtained from the model in heliocentric equatorial coordinates (kilometers).
    \item The position $\vv{R_{SB}}$ and the velocity $\vv{V_{SB}}$ of the centre of the Sun with respect to the Solar System barycentre is obtained from the model in  heliocentric equatorial coordinates (kilometers).
    \item The position $\vv{R_{EEM}}$ and the velocity $\vv{V_{EEM}}$ of the centre of the Earth with repect to the the Earth-Moon barycentre is obtained from the model in heliocentric equatorial coordinates (kilometers).
\end{enumerate}

\noindent The heliocentric position and the velocity of the centre of the Earth in equatorial coordinates is then computed as:

\begin{equation}
\begin{aligned}
    \vv{R_{EH}} = \vv{R_{EMB}} - \vv{R_{SB}} + \vv{R_{EEM}} \\
    \vv{V_{EH}} = \vv{V_{EMB}} - \vv{V_{SB}} + \vv{V_{EEM}}
\end{aligned}
\end{equation}

\noindent where $\vv{R_{EH}}$ and $\vv{V_{EH}}$ are in \SI{}{\kilo \metre} and \SI{}{\kilo \metre \per \second}, respectively.

\subsection{Heliocentric coordinates of the meteor}

Coordinates of the meteor in heliocentric equatorial coordinates can be calculated by simply adding the position of the Earth in heliocentric equatorial coordinates to the ECI coordinates of the meteor in J2000:

\begin{equation}
    \vv{M} = \vv{R_{EH}} + \frac{\vv{S'}}{1000}
\end{equation}

\noindent Care must be taken to match the units, as the ECI coordinates were given in meters, while $\vv{M}$ should be in kilometers. Both the coordinates of the meteor $\vv{M}$ and the velocity of the Earth $\vv{V_{EH}}$ have to be converted to the ecliptic reference frame by rotating them on the X axis by the negative value of the mean obliquity of the Earth at J2000, $\epsilon_{J2000} = \ang{23.4392911111}$:

\begin{equation} \label{eq:ecliptic_rotation}
\begin{aligned}
    \begin{bmatrix}
        x_{ecliptic} \\
        y_{ecliptic} \\
        z_{ecliptic}
    \end{bmatrix}
    =
    \begin{bmatrix}
        1 &                       0 &                      0 \\
        0 &  \cos (-\epsilon_{J2000}) & \sin (-\epsilon_{J2000}) \\
        0 & -\sin (-\epsilon_{J2000}) & \cos (-\epsilon_{J2000})
    \end{bmatrix}
    \begin{bmatrix}
        x_{equatorial} \\
        y_{equatorial} \\
        z_{equatorial}
    \end{bmatrix}
\end{aligned}
\end{equation}

\noindent after which $\vv{M'}$ and $\vv{V'_{EH}}$ in heliocentric ecliptic coordinates are obtained. 

The heliocentric velocity vector of the meteor is calculated by adding the geocentric velocity of the meteor to the velocity of the Earth. As $\vv{V'_{EH}}$ is in heliocentric ecliptic coordinates, we convert the geocentric velocity into an ecliptic velocity vector ($\lambda_g$ and $\beta_g$ can be computed using equations in Appendix \ref{appendix:ecliptic}):

\begin{equation} \label{eq:geocentric_eclitpic_velocity}
\begin{aligned}
    v_{gEx} = - v_g \cos \lambda_g \cos \beta_g \\
    v_{gEy} = - v_g \sin \lambda_g \cos \beta_g \\
    v_{gEz} = - v_g \sin \beta_g
\end{aligned}
\end{equation}

\noindent and add it to the velocity of the Earth around the Sun to obtain the heliocentric velocity vector $\vv{v_H}$:

\begin{equation}
    \vv{v_H} = \vv{V'_{EH}} + \frac{\vv{v_{gE}}}{1000}
\end{equation}

\noindent where $\vv{v_H}$ and $\vv{V'_{EH}}$ are \SI{}{\kilo \metre \per \second}, and $\vv{v_{gE}}$ is in \SI{}{\metre \per \second}.

\subsection{Heliocentric ecliptic radiants}

\cite{tsuchiya2017correction} have shown that low-velocity meteor showers suffer from large dispersion in geocentric equatorial coordinates due to the component of Earth's velocity. They propose calculating radiants in heliocentric ecliptic coordinates, as slower meteor showers show significantly lower dispersions in that coordinate system. For completeness, we give the equations below.

The unit heliocentric velocity vector of the meteoroid is calculated as:
\begin{equation}
    \hat{V_c} = \frac{\vv{v_H}}{|\vv{v_H}|}
\end{equation}

\noindent and the radiant in heliocentric ecliptic coordinates is then calculated as:

\begin{equation}
\begin{aligned}
    \lambda_h = \atantwo \left ( \hat{V}_{cy}, \hat{V}_{cx} \right) + \pi \\
    \beta_h = -\arcsin \hat{V}_{cz}
\end{aligned}
\end{equation}

\subsection{Keplerian orbital elements}

The solar longitude $\lambda_{\astrosun}$ can be calculated from the ecliptic heliocentric position of the Earth $\vv{R'_{EH}}$, which can be computed by rotating the equatorial heliocentric position $\vv{R_{EH}}$ using equation \ref{eq:ecliptic_rotation}.

\begin{equation}
    \lambda_{\astrosun} = \atantwo \left ( R'_{EHy}, R'_{EHx} \right) + \pi
\end{equation}

The specific orbital energy $\epsilon$ can be calculated as:

\begin{equation}
    \epsilon = \frac{|\vv{v_H}|^2}{2} - \frac{\mu_{\astrosun}}{|\vv{M'}|}
\end{equation}

\noindent where $\vv{v_H}$ is the heliocentric ecliptic velocity vector of the meteor, $\mu_{\astrosun} = \SI{1.32712440018e11}{\kilo \metre \cubed \per \second \squared}$ is the gravitational constant of the Sun, and $\vv{M'}$ is the heliocentric ecliptic position vector of the meteoroid.

The semi-major axis in AU is:

\begin{equation}
    a = \frac{-\mu_{\astrosun}}{2 \epsilon r_{AU}}
\end{equation}

\noindent where $r_{AU} = \SI{149597870.7}{\kilo \metre}$ is one astronomical unit in kilometers. Mean motion in radians per day can be calculated as:

\begin{equation}
    n = 86400 \sqrt{ \frac{ G M_{\astrosun} }{\left (1000 |a| r_{AU} \right )^3 } }
\end{equation}

\noindent where $G = \SI{6.67384e-11}{\metre \cubed \per \kilogram \per \second \squared}$ is the gravitational constant, and $M_{\astrosun} = \SI{1.98855e+30}{\kilogram}$ is the mass of the Sun. The orbital period in years is:

\begin{equation}
    T = \frac{ 2 \pi } {86400 Y_S} \sqrt{  \frac{ (r_{AU} a )^3 } {\mu_{\astrosun}}}
\end{equation}

\noindent where $Y_S = 365.256363004$ is the sidereal year in days. Next, we calculate the orbital angular momentum vector:

\begin{equation}
    \vv{h} = \vv{M'} \times \vv{v_H}
\end{equation}

\noindent the inclination is then simply:

\begin{equation}
    i = \arccos \frac{h_z}{|\vv{h}|}
\end{equation}

\noindent and the eccentricity is then the magnitude of the eccentricity vector:

\begin{equation}
\begin{aligned}
    \vv{e} = \frac{\vv{v_H} \times \vv{h}} { \mu_{\astrosun} } - \frac{ \vv{M'} }{ | \vv{M'} | } \\
    e = \left | \vv{e} \right |
\end{aligned}
\end{equation}

We follow \cite{jenniskens2011cams} on calculating the perihelion distance:

\begin{equation}
q = 
\begin{cases}
    \frac{|\vv{M'}| + \vv{e} \cdot \vv{M'} }{ 1 + |\vv{e}| }, & e = 1 \\
    a (1 - e),                                                & \text{otherwise}
\end{cases}
\end{equation}

\noindent the aphelion distance is then simply:

\begin{equation}
    Q = a (1 + e)
\end{equation}

The ascending node $\Omega$ is calculated following \cite{clark2010searching}:

\begin{equation}
\begin{aligned}
    \vv{k} = [0, 0, 1] \\
    \vv{n} = \vv{k} \times \vv{h} \\
\end{aligned}
\end{equation}

\begin{equation}
\Omega = 
\begin{cases}
    0,                                 & |\vv{n}| = 0 \\
    \atantwo \left ( n_y, n_x \right), & \text{otherwise}
\end{cases}
\end{equation}

\noindent where $\vv{n}$ is a vector pointing from the Sun to the ascending node. Please note that the ascending node loses meaning for inclinations close to $\ang{0}$, thus we keep the node at \ang{0} when the magnitude of the $\vv{n}$ vector is 0. 

If $|\vv{n}| \neq 0$, the argument of perihelion $\omega$ is calculated as:

\begin{equation}
    \omega = \arccos \frac{\vv{n} \cdot \vv{e} }{ |\vv{n}| |\vv{e}| }
\end{equation}

\noindent and if $e_z < 0$ then $\omega = 2 \pi - \omega$. If on the other hand $|\vv{n}| = 0$, then

\begin{equation}
    \omega = \arccos \frac{e_x}{|\vv{e}|}
\end{equation}

The longitude of perihelion $\varpi$ is simply:

\begin{equation}
    \varpi = \Omega + \omega
\end{equation}

True anomaly $\nu$ is calculated as:

\begin{equation}
    \nu = \arccos \frac{ \vv{e} \cdot \vv{M'} }{ |\vv{e}| |\vv{M'}| }
\end{equation}

\noindent and if $\vv{M'} \cdot \vv{v_H} < 0$ then $\nu = 2 \pi - \nu$.

The eccentric anomaly $E$ is:

\begin{equation}
    E = \atantwo \left ( \sqrt{1 - e^2} \sin \nu, e + \cos \nu \right)
\end{equation}

\noindent from which the mean anomaly $M$ can be calculated as:

\begin{equation}
    M = E - e \sin E
\end{equation}

The time in days since the last perihelion passage, reference to $TDB$, is:

\begin{equation}
    \Delta t_{\varpi} = \frac{M a^{3/2}}{k}
\end{equation}

\noindent where $k = 0.01720209895 (\textrm{AU})^{3/2} (\textrm{day})^{-1} (\textrm{solar mass})^{-1/2}$ is the Gaussian gravitational constant.

Finally, we calculate the Tisserand parameter with respect to Jupiter as:

\begin{equation}
    T_J = \frac{a_J}{a} + 2\sqrt{ (1 - e^2) \frac{ a }{ a_J } } \cos i
\end{equation}

\noindent where $a_J = 5.204267 \mathrm{AU}$ is the semi-major axis of Jupiter.

\section{Earth-centered inertial coordinates}

The Earth-centered inertial (ECI) coordinates are a Cartesian coordinate system where the X-Y plane coincides with the equatorial plane of the Earth, and the X axis passes through the equinox of the given epoch. The Z axis passes through the Earth's North pole. As the coordinate system is permanently fixed to the celestial sphere, a fixed point on the surface of the Earth will have changing coordinates in time. As we assume that the observations are given in the epoch of date, we keep the ECI coordinates in the epoch of date as well.

Let the distance from the centre of the Earth to the position given by geographical coordinates in the WGS84 system be calculated as follows:
    
\begin{equation}
    N = \frac{r_e}{\sqrt{1 - e_e^2 \sin^2 \varphi }}
\end{equation}

\noindent where $r_e$ is the equatorial radius of the Earth as defined by the WGS84 system, $r_e = \SI{6378137.0}{\metre}$, and $e_e$ is the equatorial ellipticity of an oblate Earth:

\begin{equation}
    e_e = \sqrt{1 - \frac{r_p^2}{r_e^2}}
\end{equation}

\noindent where $r_p$ is the polar radius of the Earth, $r_p = \SI{6356752.314245}{\metre}$. The polar ellipticity is:

\begin{equation}
    e_p = \sqrt{\frac{r_e^2 - r_p^2}{r_p^2}}
\end{equation}

\subsection{Converting geographical coordinates to ECI} \label{appendix:geo2eci}

Let $\varphi$ be the geodetic latitude, $\lambda$ the longitude, $h$ the height above a WGS84 model for Earth, and $\theta'$ the apparent local sidereal time (LST). Note that the height is not the same as the Mean Sea Level (MSL) height reported by Google Earth and some GPS devices. If the MSL height is used, it has to be first converted to WGS84 height \citep{pavlis2012development}. The apparent LST $\theta'$ can be calculated using the procedure described in \cite{meeus1998astronomical} page 88 and \cite{clark2010searching}, see Appendix \ref{appendix:lst} for equations.

First, the coordinates are transformed into Earth-Centred Earth-Fixed (ECEF) coordinates:

\begin{equation}
\begin{aligned}
    x_{ECEF} = (N + h) \cos \varphi \cos \lambda \\
    y_{ECEF} = (N + h) \cos \varphi \sin \lambda \\
    z_{ECEF} = \left( (1 - e_e^2)N + h \right ) \sin \varphi
\end{aligned}
\end{equation}

The radius of the Earth at the given geodetic latitude is then:

\begin{equation}
    R_h = \sqrt{x_{ECEF}^2 + y_{ECEF}^2 + z_{ECEF}^2}
\end{equation}

Using the geocentric latitude $\varphi'$ :

\begin{equation}
    \varphi' = \atantwo \left (z_{ECEF}, \sqrt{x_{ECEF}^2 + y_{ECEF}^2} \right )
\end{equation}

The ECI coordinates in the epoch of date are then calculated as:

\begin{equation}
\begin{aligned}
    x_{ECI} = R_h \cos \varphi' \cos \theta' \\
    y_{ECI} = R_h \cos \varphi' \sin \theta' \\
    z_{ECI} = R_h \sin \varphi'
\end{aligned}
\end{equation}

\subsection{ECI to geographical coordinates} \label{appendix:eci2geo}

Given the apparent sidereal time at Greenwich $\theta_0'$ (see equation \ref{eq:gst}), the longitude can be calculated as:

\begin{equation}
    \lambda = \atantwo \left ( y_{ECI}, x_{ECI} \right) - \theta_0'
\end{equation}

The geodetic latitude $\varphi$ is calculated as:

\begin{equation}
\begin{aligned}
    p = \sqrt{x_{ECI}^2 + y_{ECI}^2} \\
    \vartheta = \atantwo \left ( z_{ECI} r_e, p r_p \right ) \\
    \varphi = \atantwo \left ( z_{ECI} + e_p^2 r_p \sin^3 \vartheta, p - e_e^2 r_e \cos^3 \vartheta \right )
\end{aligned}
\end{equation}

Care must be taken when calculating the height near exact poles due to numerical instabilities. If the coordinates are near the poles, and we take this as being within \SI{1}{\kilo \metre} from the poles, which can be determined by testing if both conditions $|x_{ECI}| < 1000$ and $|y_{ECI}| < 1000$ are true, the height is calculated as:

\begin{equation}
    h = |z_{ECI}| - r_p
\end{equation}

\noindent otherwise, the height above a WGS84 ellipsoid is calculated as:

\begin{equation}
\begin{aligned}
    N = \frac{r_e}{\sqrt{1 - e_e^2 \sin^2 \varphi}} \\
    h = \frac{p}{\cos \varphi} - N
\end{aligned}
\end{equation}

This height is given in the WGS84 convention, if the height above mean sea level (MSL) is desired, a correction described in \cite{pavlis2012development} has to be applied.

\section{Local apparent sidereal time} \label{appendix:lst}

First, we calculate the nutation components $\Delta \psi$ and $\Delta \epsilon$ in equation \ref{eq:nutation_components} as given in \cite{meeus1998astronomical}, chapter 22. We use the set of equations which give around \SI{0.5}{\arcsecond} precision, which we deem sufficient for needs of meteoroid orbits. The dynamical time $TDB$ is used. $\Omega$ is the longitude of the ascending node of the Moon's mean orbit on the ecliptic measured from the mean equinox of the date, $L$ is the mean longitude of the Sun, and $L'$ is the mean longitude of the Moon. The values are in degrees.

\begin{equation}
\begin{aligned}
    T = \frac{TDB - 2451545}{36525} \\
    \Omega = 125.04452 - 1934.136261 T \\
    L = 280.4665 + 36000.7698 T \\
    L' = 218.3165 + 481267.8813 T
\end{aligned}
\end{equation}

The nutation in longitude $\Delta \psi$ and the nutation in obliquity $\Delta \epsilon$ are calculated in arc seconds as:

\begin{equation} \label{eq:nutation_components}
\begin{aligned}
    \Delta \psi = -17.2 \sin \Omega - 1.32 \sin 2 L - 0.23 \sin 2 L' + 0.21 \sin 2 \Omega \\
    \Delta \epsilon = 9.2 \cos \Omega + 0.57 \cos 2 L + 0.1 \cos 2 L' - 0.09 \cos 2 \Omega
\end{aligned}
\end{equation}

Next, we calculate the mean sidereal time of the Earth (Greenwich Sidereal Time) in degrees. Note that the time used here is not dynamical.

\begin{equation} 
\begin{aligned}
    t = \frac{JD - 2451545}{36525} \\
    \theta_0 = 280.46061837 + 360.98564736629 (JD - 2451545) \\
                + 0.000387933 t^2 - \frac{t^3}{38710000}
\end{aligned}
\end{equation}

The mean obliquity of the Earth in arc seconds $\epsilon_0$ is calculated using $U$, which is the time measured in units of 10000 Julian years from J2000 (note that the dynamical time is used):

\begin{equation} \label{eq:mean_obliquity}
\begin{aligned}
    U = \frac{TDB - 2451545}{3652500} \\
    \epsilon_0 = 84381.448 - 4680.93 U \\
                            - 1.55 U^2 \\
                            + 1999.25 U^3 \\
                            - 51.38 U^4 \\
                            - 249.67 U^5 \\
                            - 39.05 U^6 \\
                            + 7.12 U^7 \\
                            + 27.87 U^8 \\
                            + 5.79 U^9 \\
                            + 2.45 U^{10}
\end{aligned}
\end{equation}

The apparent sidereal time at Greenwich in degrees is calculated as:

\begin{equation} \label{eq:gst}
\begin{aligned}
    \theta_0' = \theta_0 + \frac{\Delta \psi}{3600} \cos \frac{\epsilon_0 + \Delta \epsilon}{3600}
\end{aligned}
\end{equation}

\noindent After converting to radians, care must be taken to wrap the computed value inside the $[0, 2 \pi]$ range using modulus operator:

\begin{equation}
    \theta_0' = \theta_0' \bmod 2 \pi
\end{equation}

Finally, the apparent local sidereal time $\theta'$ can be calculated as:

\begin{equation}
    \theta' = (\theta_0' + \lambda + 2 \pi )\bmod 2 \pi
\end{equation}

\noindent where $\lambda$ is the geodetic longitude of the observer.

\section{Horizontal to equatorial coordinate conversion} \label{appendix:altaz2radec}
    
Right ascension $\alpha$ and declination $\delta$ are calculated from azimuth $A$, altitude $a$, Julian date $JD$, and geographical coordinates of the observer, longitude $\lambda$ and latitude $\varphi$, as:

\begin{equation}
\begin{aligned}
    H = \atantwo \left ( -\sin A, \tan a \cos \varphi - \cos A \sin \varphi \right) \\
    \alpha = \theta' - H \\
    \delta = \arcsin \left ( \sin \varphi \sin a + \cos \varphi \cos a \cos A \right )
\end{aligned}
\end{equation}

\noindent where $H$ is the local hour angle and $\theta'$ is the apparent local sidereal time (see Appendix \ref{appendix:lst}).

\section{Equatorial to horizontal coordinate conversion} \label{appendix:radec2altaz}

The azimuth $A$ and altitude $a$ are calculated from right ascension $\alpha$, declination $\delta$ Julian date $JD$, and geographical coordinates of the observer, longitude $\lambda$ and latitude $\varphi$, as:

\begin{equation}
\begin{aligned}
    H = \theta' - \alpha \\
    A = \pi + \atantwo \left ( \sin H, \cos H \sin \varphi - \tan \delta \cos \varphi \right ) \\
    a = \arcsin\left ( \sin \varphi \sin \delta + \cos \varphi \cos \delta \cos H \right)
\end{aligned}
\end{equation}

\noindent where $H$ is the local hour angle and $\theta'$ is the apparent local sidereal time (see \ref{appendix:lst}).

\section{Precessing equatorial coordinates} \label{appendix:precession_eq}

We follow the rigorous method of \cite{meeus1998astronomical}, pages 134 - 135, for precessing the right ascension $\alpha$ and declination $\delta$ from epoch $JD_0$ to epoch $JD$. The beginning of each epoch is defined by their respective Julian dates. Please note that $\zeta$, $z$ and $\theta$ are given in degrees.

\begin{equation}
\begin{aligned}
    T = \frac{JD_0 - 2451545}{36525} \\
    t = \frac{JD - JD_0}{36525} \\
    \zeta  = \frac{1}{3600} \left[ (2306.2181 + 1.39656 T - 0.000139 T^2) t \right. \\
            \left. + (0.30188 - 0.000344 T) t^2 + 0.017998 t^3 \right] \\
    z      = \frac{1}{3600} \left[(2306.2181 + 1.39656 T - 0.000139 T^2) t \right. \\
            \left. + (1.09468 + 0.000066 T) t^2 + 0.018203 t^3 \right] \\
    \theta = \frac{1}{3600} \left[(2004.3109 - 0.85330 T - 0.000217 T^2) t \right. \\
            \left. - (0.42665 + 0.000217 T) t^2 - 0.041833 t^3 \right] \\
    A = \cos \delta \sin(\alpha + \zeta) \\
    B = \cos \theta \cos \delta \cos(\alpha + \zeta) - \sin \theta \sin \delta \\
    C = \sin \theta \cos \delta \cos(\alpha + \zeta) + \cos \theta \sin \delta \\
    \alpha' = \atantwo(A, B) + z \\
    \delta' = \arcsin C 
\end{aligned}
\end{equation}

\noindent where $\alpha'$ and $\delta'$ are precessed coordinates. If the declination is close to the celestial poles (which we define this as less than $\ang{0.5}$ from the poles), it is calculated differently due to numerical instabilities. If $(\ang{90} - |\delta|) < \ang{0.5}$ is true, the declination should be calculated as:

\begin{equation}
    \delta' = \arccos \sqrt{A^2 + B^2}
\end{equation}

\section{Ecliptic coordinates} \label{appendix:ecliptic}

The geocentric right ascension $\alpha_g$ and declination $\delta_g$ at the given epoch (the epoch defined by a Julian date $JD$, usually at J2000, thus $JD$ = 2451545) can be converted to geocentric ecliptic longitude $\lambda_g$ and latitude $\beta_g$ with the procedure described below. First, a precise obliquity of the Earth at the $JD$ of the epoch has to be calculated; $\Delta \epsilon$ can be calculated using equation \ref{eq:nutation_components} and the mean obliquity $\epsilon_0$ using equation \ref{eq:mean_obliquity}. The true obliquity of the Earth in degrees is then simply:
    
\begin{equation}
    \epsilon = \frac{\epsilon_0 + \Delta \epsilon}{3600}
\end{equation}

\noindent The ecliptic longitude and latitude are then:

\begin{equation}
\begin{aligned}
    \lambda_g = \atantwo \left ( \sin \epsilon \sin \delta_g + \sin \alpha_g \cos \delta_g \cos \epsilon, \cos \alpha_g \cos \delta_g \right ) \\
    \beta_g = \arcsin \left ( \cos \epsilon \sin \delta_g - \sin \alpha_g \cos \delta_g \sin \epsilon \right)
\end{aligned}
\end{equation}


\bsp	
\label{lastpage}
\end{document}